\documentclass{aa}
\usepackage{pdflscape} % or {lscape}
\usepackage{natbib}
\usepackage{wasysym}
\usepackage[varg]{txfonts}
\usepackage{graphicx}
\usepackage[draft]{hyperref}
\bibpunct{(}{)}{;}{a}{}{,} % to follow the A&A style
\usepackage{longtable}

\usepackage{afterpage}
\usepackage{ulem}

\usepackage[usenames,dvipsnames]{color}
\definecolor{mygreen}{rgb}{0,0.5,0}
\definecolor{myorange}{rgb}{0.5,0.5,0}
\definecolor{myred}{rgb}{0.5,0,0}

\def\ms{\hbox{\,m\,s$^{-1}$}}         %m.s -1
\def\cms{\hbox{\,cm\,s$^{-1}$}}       %cm.s -1
\def\m2s2{\hbox{\,m$^{2}$\,s$^{-2}$}} %m2.s -2
\def\kms{\hbox{\,km\,s$^{-1}$}}       %km.s -1
\def\logrhk{$\log$(R$^{\prime}_{HK}$)}

\begin{document}

\title{Measuring precise radial velocities and cross-correlation function line-profile variations using a Skew Normal density\thanks{Based on observations collected at the La Silla Parana Observatory,
ESO (Chile), with the HARPS spectrograph at the 3.6-m telescope.}}

\author{U. Simola \inst{1}
	    \thanks{\email{umberto.simola@helsinki.fi}}
	    \and X. Dumusque\inst{2}
	    \thanks{Branco Weiss Fellow--Society in Science (url: \url{http://www.society-in-science.org})}    
	    \and Jessi Cisewski-Kehe\inst{3}
	    }

\institute{Department of Mathematics and Statistics, University of Helsinki, Helsinki, Finland
	      \and Observatoire de Gen\`eve, Universit\'e de Gen\`eve, 51 ch. des Maillettes, CH-1290 Versoix, Switzerland 
	      \and Department of Statistics and Data Science, Yale University, New Haven, CT, USA
	      }

\date{Received XXX; accepted XXX}

\abstract
{Stellar activity is one of the primary limitations to the detection of low-mass exoplanets using the radial-velocity (RV) technique. 
Stellar activity can be probed by measuring time-dependent variations in the shape of the cross-correlation function (CCF). It is therefore 
critical to measure with high-precision these shape variations to de-correlate the signal of an exoplanet from spurious RV signals caused 
by stellar activity.}
{We propose to estimate the variations in shape of the CCF by fitting a Skew Normal (SN) density which, unlike the commonly employed Normal density, includes a skewness parameter to capture the asymmetry of the CCF induced by stellar activity and the convective blueshift.
}
{The performances of the proposed method are compared to the commonly employed Normal density using both simulations and real observations, with different levels of activity and signal-to-noise ratio.}
{When considering real observations, the correlation between the RV and the asymmetry of the CCF and between the RV and the width of the CCF are stronger when using the parameters estimated with the SN density rather than the ones obtained with the commonly employed Normal density. 
In particular the strongest correlations have been obtained when using the mean of the SN as an estimate for the RV. 
This suggests that the CCF parameters estimated using a SN density are more sensitive to stellar activity, which can be helpful 
when estimating stellar rotational periods and when characterizing stellar activity signals.
Using the proposed SN approach, the uncertainties estimated on the RV defined as the median of the SN are on average $10\%$ smaller than the uncertainties calculated on the mean of the Normal. 
The uncertainties estimated on the asymmetry parameter of the SN are on average $15\%$ smaller than the uncertainties measured on the Bisector Inverse Slope Span (BIS SPAN), which is the commonly used parameter to evaluate the asymmetry of the CCF. 
We also propose a new model to account for stellar activity when fitting a planetary signal to RV data.
Based on simple simulations, we were able to demonstrate that this new model improves the planetary detection limits by 12\% 
compared to the model commonly used to account for stellar activity.
}
{The SN density is a better model than the Normal density for characterizing the CCF since the correlations used to probe stellar activity are stronger and the uncertainties of the RV estimate and the asymmetry of the CCF are both smaller.
}


\titlerunning{Fitting a SN distribution to CCF}
\authorrunning{U. Simola, X. Dumusque and J. Cisewski}
\maketitle

%-----------------------------------------------------------------------------------------------------------------------------------------------
\section{Introduction} \label{intro}

%%%%%%%%%%%%%%%%%%%%%%%%%%%%%%%%%%
%\subsection{Goal of RV analysis}

When working with radial-velocities data (RVs), one of the main limitations to the detection of low-mass exoplanets is no longer the precision of the instruments used, but the different sources of variability induced by the stars \citep[e.g.][]{Feng:2017aa, Dumusque:2017aa, Rajpaul-2015, Robertson-2014}. 
Stellar oscillations, granulation phenomena, and stellar activity can all induce apparent RV signals that are above the meter-per-second (\ms) precision \citep[e.g.][]{Saar-1997b, Queloz-2001, Desort-2007, Dumusque-2011a, Dumusque-2016a} reached by the best high-resolution spectrographs \citep[HARPS, HARPS-N,][]{Mayor-2003,Cosentino-2012}.
It is therefore mandatory to better understand stellar signals and to develop methods to correct for them, if in the near future we want to detect or confirm an Earth-twin planet using the RV technique. This is even more true now that instruments like the Echelle SPectrograph for Rocky Exoplanet and Stable Spectroscopic Observations (ESPRESSO) \citep{Pepe-2014} and the EXtreme PREcision Spectrometer (EXPRES) \citep{fischer2016state} should reach the precision and stability to detect such signals. However, if solutions are not found to mitigate the impact of stellar activity, the detection or confirmation of potential Earth-twins will be extremely challenging and false detections could plague the field.

%%%%%%%%%%%%%%%%%%%%%%%%%%%%%%%%%%
%\subsection{Stellar activity effect on CCF, Normal fit plus FWHM and BIS SPAN indicators}

One of the most challenging stellar signals to characterize and to correct for are the signals induced by stellar activity. 
Stellar activity is responsible for creating magnetic regions on the surface of stars, and those regions change locally the temperature and the convection, which can induce spurious RVs variations \citep[e.g.][]{Meunier-2010a, Dumusque-2014b, Borgniet-2015}. 
In theory, it should be easy to differentiate between the Doppler-shift induced by a planet, which shifts the entire stellar spectrum, and the effect of stellar activity, which modifies the shape of spectral lines and by doing so creates a spurious shift of the stellar spectrum \citep{Saar-1997b,Hatzes-2002,Kurster2003,Lindegren-2003,Desort-2007,Lagrange-2010,Meunier-2010a,Dumusque-2014b}. 
However, on quiet GKM dwarfs, the main targets for precise RVs measurements, stellar activity can induce signals of a few \ms. 
This corresponds physically to variations smaller than 1/100th of a pixel on the detector, making changes in the shape of the spectral lines challenging to detect.
In order to measure such tiny variations, a common approach is to average the information of all the lines in the spectrum by cross correlating the stellar spectrum with a synthetic or an observed stellar template \citep[][]{Baranne-1996,Pepe-2002a, Anglada-Escude-2012}. The result of this operation gives us the cross-correlation function (CCF).  
%The CCF gives the spectrum's cross-correlation with the template as the template is shifted according to different RVs.
%
To measure the Doppler-shift between different spectra, and therefore to retrieve the RVs of a star as a function of time, the variations of the CCF barycenter are calculated. 
The barycenter is generally estimated by the mean of  a Normal density shape fit to the CCF. 
Variations in line shape between different spectra, which indicate the presence of signals induced by stellar activity, are measured by analyzing different parameters of models fit to the CCF. Usually the width of the CCF is estimated using the full-width half-maximum (FWHM) of the fitted Normal density and its asymmetry by calculating the CCF bisector and measuring the bisector inverse slope span \citep[BIS SPAN,][]{Queloz-2001}.

%%%%%%%%%%%%%%%%%%%%%%%%%%%%%%%%%%
%\subsection{Why FWHM and BIS SPAN important?}

If a spurious RV signal is induced by activity, generally a strong correlation will be observed between the RV and chromospheric activity indicators like \logrhk\,or H-$\alpha$ \citep{Boisse-2009,Dumusque-2012,Robertson-2014}, but also between the RV and the FWHM of the CCF or its BIS SPAN \citep[][]{Queloz-2001,Boisse-2009,Queloz-2009,Dumusque-2016a}. 
Therefore a common strategy when fitting a Keplerian signal to a set of estimated RVs when searching for a planet is to include linear terms in the model to account for activity, such as the \logrhk, the FWHM, and the BIS SPAN \citep{Dumusque:2017aa,Feng:2017aa}.
It is also common to add a Gaussian process to the model to account for the correlated noise induced by stellar activity. The hyperparameters of the Gaussian process can be trained on different activity indicators \citep{Haywood-2014,Rajpaul-2015} or directly on the RVs \citep{Faria-2016a}. It is therefore essential for mitigating stellar activity to obtain activity indicators that are the most correlated with the RVs but also for which we can obtain the best precision.

%%%%%%%%%%%%%%%%%%%%%%%%%%%%%%%%%%
%\subsection{Figueira indicators of stellar activity + other}
Several indicators have been developed that can be more sensitive to line asymmetry than the BIS SPAN. In \citet{Boisse-2011}, the authors developed $V_{span}$, which is the difference between the RV measured by fitting a Normal density to the upper and the lower parts of the CCF. This CCF asymmetry parameter is shown to be more sensitive than the BIS SPAN at low signal-to-noise ratio (S/N).
\citet{Figueira-2013} studied the use of new indicators, BIS-, BIS+, bi-Gauss and $V_{asy}$. The authors were able to show that when using bi-Gauss, the amplitude in asymmetry is 30\% larger than when using BIS SPAN, therefore allowing the detection of lower levels of activity. They also demonstrated that $V_{asy}$ seems to be a better indicator of line asymmetry at high S/N, as its correlation with RV is significantly stronger than any other correlation between the previously proposed asymmetry indicators and RV.

%%%%%%%%%%%%%%%%%%%%%%%%%%%%%%%%%%
%\subsection{Why using a SN density?}
In all the methods described above, except bi-Gauss, the RV and the FWHM are derived using a Normal density fitted to the CCF, and the asymmetry is estimated using a separate approach. 
In this paper we propose to use a Skew Normal (SN) density to estimate with a single fit of the CCF the RV, the FWHM and the asymmetry of the CCF, as this function includes a skewness parameter \citep[][]{Azzalini1985}. 

%In addition, we know that for solar-type stars and cooler dwarfs, the bisector of the CCF has a "C"-shape due to convective blueshift \citep{Dravins-1981, Gray-2009}. 
%Therefore, fitting the CCF using a model that naturally includes an asymmetry, like the SN density, should give in principle more precise results.

%%%%%%%%%%%%%%%%%%%%%%%%%%%%%%%%%%
%\subsection{Outline of paper}
The paper is organized as follows. In Sec.~\ref{sec:2} we introduce the SN density, describe its applicability for modelling the CCF and study how the SN parameters relate to the mean of the Normal density, the FWHM, and the BIS SPAN of the CCF. 
In Sec.~\ref{sec:31} we propose a linear model to correct for stellar activity signals in RVs, which extends the linear models previously proposed for this purpose \citep[e.g.][]{Dumusque:2017aa,Feng:2017aa}. 
In Sec.~\ref{sec:soap} the performance of the SN fit to the CCF is investigated using simulations coming from the Spot Oscillation And Planet 2.0 code \citep[SOAP 2.0,][]{Dumusque-2014b}, followed by an analysis of real observations in Sec.~\ref{sec:4}.
In Sec.~\ref{sec:5} error bars are computed for the different estimated CCF parameters. Finally the discussion of the results and the conclusions are presented, respectively, in Secs.~\ref{sec:discu} and ~\ref{sec:conclu}.

%-----------------------------------------------------------------------------------------------------------------------------------------------
\section{The Skew Normal distribution} \label{sec:2}

\begin{figure}[t]
\begin{center}
\includegraphics[height = 2.3in]{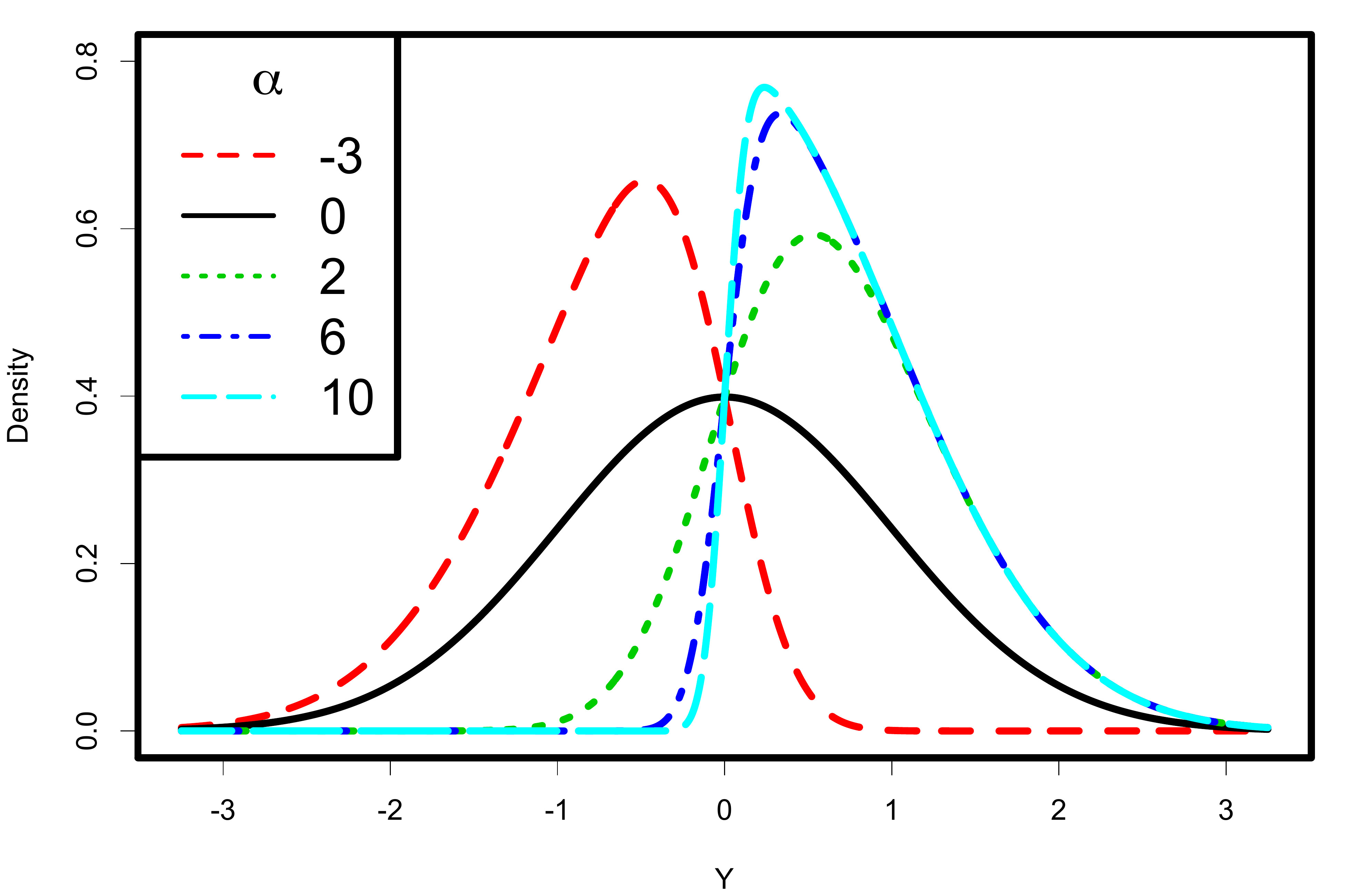} 
   \caption{Density function of a random variable $Y$ following a SN distribution $SN(\xi, \omega, \alpha)$ with location parameter $\xi = 0$, scale parameter $\omega = 1$ and different values of the skewness parameter $\alpha$ indicated by different colors and line types. Note that the solid black line has an $\alpha = 0$ making it a Normal distribution.}
   \label{fig:SN.plot}
\end{center}
\end{figure}

The Skew Normal (SN) distribution is a class of probability distributions which includes the Normal distribution as a special case \citep{Azzalini1985}. The SN distribution has, in addition to a location and a scale parameter analogous to the Normal distribution's mean and standard deviation, a third parameter which describes the skewness (i.e. the asymmetry) of the distribution. Considering a random variable $Y\in \mathbb R$ (where $\mathbb R$ is the real line) which follows a SN distribution with location parameter $\xi \in \mathbb R$, scale parameter $\omega \in \mathbb R^{+}$ (i.e., the positive real line), and skewness parameter $\alpha \in \mathbb R$, its density at some value $y\in Y$ can be written as 
\begin{equation} \label{def:snd_gen}
SN(y;\xi, \omega, \alpha) = \frac{2}{\omega} \phi\left(\frac{y-\xi}{\omega}\right) \Phi\left(\frac{\alpha(y-\xi)}{\omega}\right),
\end{equation}
where $\phi$ and $\Phi$ are, respectively, the density function and the distribution function of a standard Normal distribution\footnote{A standard Normal distribution is a Normal distribution with a mean of 0 and a standard deviation of 1.}.
The skewness parameter $\alpha$ quantifies the asymmetry of the SN. 
Examples of SN densities under different skewness parameter values and the same location and scale parameters ($\xi = 0$ and $\omega = 1$) are displayed in Fig.~\ref{fig:SN.plot}.  A usual Normal distribution is the special case of the SN distribution when the skewness parameter $\alpha$ is equal to zero\footnote{This can be seen from Eq.~\eqref{def:snd_gen}. If $\alpha = 0$ then $\Phi\left(\frac{\alpha(y-\xi)}{\omega}\right) = \Phi(0) = 0.5$ and therefore $SN(y;\xi, \omega, 0) = \frac{1}{\omega} \phi\left(\frac{y-\xi}{\omega}\right)$ which is the density of a Normal distribution. Note that $\Phi(0) = 0.5$ because $\Phi(0)$ is the the probability that a standard Normal random variable is less than or equal than 0.}.
For reasons related to the interpretation of the parameters in Eq.~\eqref{def:snd_gen} and computational issues with estimating $\alpha$ near 0, a different parametrization is used in this work, which is referred to as the \textit{centered parametrization} (CP).  This CP is much closer to the parametrization of a Normal distribution, as it uses a mean parameter $\mu$, a variance parameter $\sigma^2$ and a skewness parameter $\gamma$. In order to define the CP, we need to express the CP parameters $(\mu, \sigma^2, \gamma)$ as a function of $(\xi, \omega^2, \alpha)$. This can be done using the following relations:
\begin{equation} \label{eq:snd_cp}
\mu = \xi + \omega \beta, \quad \sigma^{2} = \omega^{2}(1-\beta^2), \quad \gamma = \frac{1}{2}(4-\pi) \beta^{3}\left(1-\beta^2\right)^{-3/2},
\end{equation}
where $\beta = \sqrt{\frac{2}{\pi}}\left(\frac{\alpha}{\sqrt{1+\alpha^2}}\right)$ \citep[e.g.][]{Arellano-2010}.

By using Eq.~\eqref{eq:snd_cp}, the new set of parameters $(\mu, \sigma^2, \gamma)$ provides a clearer interpretation of the behavior of the SN distribution. For the $\alpha$ values used in Fig.~\ref{fig:SN.plot}, the corresponding values of ($\mu$, $\sigma^2$, $\gamma$) are displayed in Table~\ref{tab:cp_values}.  In particular, $\mu$ and $\sigma^2$ are the actual mean and variance of the distribution, rather than simply a location and scale parameter, and $\gamma$ provides a measure of the skewness of the SN. 
Along with the mean of the SN, we consider the median of the distribution as a measure of its barycenter.  See Table~\ref{tab:cp_values} for the medians of the SN densities displayed in Fig.~\ref{fig:SN.plot}.

%The median of the SN, and in general the median of an absolute continuous random variable, is defined as the value $m$ such that\footnote{We recall that when using a symmetric distribution such as the Normal distribution, the mean and the median are equivalent.}:
%%
%\begin{equation} \label{eq:snmed}
%\int_{-\infty}^{m} SN(y;\xi, \omega, \alpha) = \frac{1}{2}.
%\end{equation}
%

%% Requires the booktabs if the memoir class is not being used
\begin{table}[htbp]
\begin{center}
   \caption{CP values $(\mu, \sigma^2, \gamma)$ along with the median corresponding to the $\alpha$ values shown in Fig.~\ref{fig:SN.plot}, with location parameter $\xi = 0$ and scale parameter $\omega = 1$. Values are rounded to three decimal places.}
   \label{tab:cp_values}
   \begin{tabular}{|ccccc|} % Column formatting, @{} suppresses leading/trailing space
\hline
$\alpha$ & $\mu$ & $\sigma^2$ & $\gamma$  & Median \\
\hline
 -3 	&	 -0.757	&	 0.427	&	 -0.667  	& 	-0.672\\
0	&	 0.000 	&	1.000	&	 0.000 	& 	0.000\\
2	&	 0.714	&	 0.491	&	 0.454 	& 	0.655\\
6	&	 0.787	&	 0.381	&	 0.891 	& 	0.674\\
10	&	 0.794	&	 0.370	&	 0.956 	& 	0.674\\
\hline
   \end{tabular}

\end{center}
\end{table}
Further details about the parametrization from Eq.~\eqref{def:snd_gen}, called the \textit{Direct Parametrization} (DP), the CP, and general statistical properties of the SN can be found in \cite{Azzalini2014}.

%-----------------------------------------------------------------------------------------------------------------------------------------------
\subsection{Fitting the Skew Normal density to the CCF} \label{sec:3}

%The CCF represents the average shape of spectral lines and is expressed in flux as a function of radial-velocity.
To fit the CCF using a SN density shape, we use a least-squares algorithm and the following model:
\begin{eqnarray} \label{eq:3}
f_{CCF}(x_i) = \mathrm{C} - \mathrm{A} \times SN(x_i;\mu, \sigma^2, \gamma), \quad i = 1, \ldots, n
\end{eqnarray}
where C is an unknown offset for the continuum of the CCF, A is the unknown amplitude of the CCF, sometimes referred to as the CCF contrast, and $\mu$, $\sigma^2$ and $\gamma$ are, respectively, the mean, the variance, and the skewness of the SN as defined above.
The values $x_1, \ldots, x_n$ are the different values of the x-axis of the CCF, generally in velocity units (e.g. \ms).

%%%Since the CCF has an asymmetry due the convective blueshift, the SN density should in principle better catch this aspect, together with other changes in asymmetry, with respect to fitting a Normal density. 
%%%To initially check this intuition, we compared the CCF residuals after fitting a Normal and a SN density for 2 stars. The first star is Alpha Centauri B, whose CCF's have high signal-to-noise ratio (S/N). The second star is Corot-7, whose CCF's have low S/N. Fig.~\ref{fig:Residual.comparison} shows that the SN seems to be a slightly better model to explain the shape of the CCF.%, in particular as the S/N decreases.
%%%%
%%%\begin{figure*}[htbp]
%%%   \centering
%%%\includegraphics[height = 2.5in]{[1]HD128621Residuals.pdf} 
%%%\includegraphics[height = 2.5in]{[1]LRa01_E2Residuals.pdf} 
%%%   \caption{Comparison between the Normal (black circles) and the SN (red pluses) residuals using CCF's from the star Alpha Centauri B (left) and Corot-7 (right). When looking at the residuals corresponding to the center of the CCF, the SN fit leads to slightly better results for both stars.} %Moreover, as the S/N decreases, the SN density shows smaller residuals respect the Normal ones.}
%%%    \label{fig:Residual.comparison}
%%%\end{figure*}
%%%%

When using a Normal density shape model for the CCF, the estimated mean is used as the estimated RV and the FWHM\footnote{FWHM$=2\sqrt{2\ln2}\,\sigma$ with standard deviation $\sigma$} is used to quantify the width of the CCF.
Because the Normal density is symmetric, the skewness is not defined and therefore a separate approach is necessary to estimate the skewness of the CCF.
An estimated skewness parameter is generally obtained by calculating the BIS SPAN of the CCF \citep[see Sec.~\ref{intro}, and e.g.][]{Queloz-2001}. 

With the proposed SN approach, we propose two estimators of the RV: the mean and median of the SN model fit (referred to as SN mean RV and SN median RV, respectively), and present advantages and limitations for both of these choices in Sec.~\ref{sec:4} and Sec.~\ref{sec:5}. 
The width of the SN, SN FWHM, is defined in the same way as for the Normal density\footnote{Note that SN FWHM does not correspond to the width of the SN density at half maximum like in the Normal case.}, and finally the skewness of the CCF is estimated by the $\gamma$ parameter.

To evaluate the strength of the correlation between the estimated RVs and the different stellar activity indicators, we calculated the Pearson correlation coefficient, $R$, which in its general form is defined as:
\begin{equation}
R (x,y)= \frac{\text{cov}(x,y)}{\sigma(x)\sigma(y)},
\label{eq:Pearson:corr}
\end{equation}
where $x$ and $y$ are two quantitative variables, $\text{cov}(x,y)$ indicates the covariance between $x$ and $y$, and $\sigma(x)$ and $\sigma(y)$ represent their standard deviations.  A $p$-value for the statistical test with null hypothesis $H_{0}: R=0$ is also generally provided.

%-----------------------------------------------------------------------------------------------------------------------------------------------
\section{Radial Velocity correction for stellar activity} \label{sec:31}

Exoplanets produce a Doppler shift of the entire stellar spectrum. On the contrary stellar activity and, in particular, the presence of active regions on the stellar photosphere, do not produce blueshifts or redshifts of the entire stellar spectrum but can create spurious RV signals by modifying the shape of spectral lines.
To track these variations in the shape of the spectral lines, a common approach consists in using the FWHM, the BIS SPAN, or other indicators such as those introduced in \citet{Boisse-2011} or \citet{Figueira-2013}, which provide information on the width and asymmetry of the CCF. A strong correlation between the estimated RVs and one or more of these parameters provides an indication that stellar activity signals may be affecting the measurements.

When fitting for planetary signals in RV data, it is common to include linear dependencies with the BIS SPAN and the FWHM to take into account the signal induced by stellar activity \citep[e.g.][]{Dumusque:2017aa,Feng:2017aa}.
%Some people also include the \logrhk, however, \citet{Feng-2018} show that this is probably not a good choice as \logrhk and RV are not well correlated.
We propose to add additional parameters in the model to correct for stellar activity: 
(i) an amplitude parameter A of the CCF (referred to as the CCF contrast) and (ii) an interaction term for $\gamma$ and SN FWHM (or the BIS SPAN and the FWHM in the Normal case). The stellar activity correction we propose can therefore be written as:
\begin{equation}
RV_{\text{activity}}= \beta_{0} + \beta_{1} A + \beta_{2} \gamma + \beta_{3} \text{SN FWHM} + \beta_{4} (\gamma  \text{SN FWHM})+\epsilon,
\label{eq:RV:correction}
\end{equation}
where $\beta_{0}$ is the intercept and $\epsilon$ is the random error with mean equal to $0$ and covariance matrix equal to $\sigma^{2}I$ ($I$ defined as the identity matrix). 
The contrast parameter $A$ accounts for the presence of a spot on the stellar surface, which produces a change in the amplitude of the CCF, in addition to changes in asymmetry or width \citep[see e.g. Fig. 2 in ][]{Dumusque-2014b}.
The benefits of including a variable that quantifies the interaction between $\gamma$ and SN FWHM (or BIS SPAN and FWHM) will be better understood through the results of the examples presented in Sec.~\ref{sec:soap}. 
This interaction term can account for possible interactions between SN FWHM (or FWHM) and $\gamma$ (or BIS SPAN), meaning that each  variables' association with the response, $RV_{\text{activity}}$, depends also on the other variable.
%while the association between in this case BIS SPAN and FWHM (or $\gamma$ and SN FWHM) means that the values of one variable relate to the values of the other (since in this case we have two quantitative variables we talk about correlation), with the term interaction we mean that the effect that one variable has on the RVs is not constant. In particular the effect differs at different values of the other values. %As a consequence of this, if two variables are interacting the may or may not be associated.

The proposed model is analyzed using statistical tests on the parameters $\beta_{0}$, $\beta_{1}$, $\beta_{2}$, $\beta_{3}$ and $\beta_{4}$ where the null hypothesis is $H_{0}: \beta_{i}=0$, for $i=0,\dots,4$. The significance level for the tests are set at $0.05$. The coefficient of determination, $R^2$, is used to assess how well the proposed linear combination of variables accounts for the variability of $RV_{\text{activity}}$. 

The proposed function defined in Eq.~\eqref{eq:RV:correction} is the result of statistical and astronomical considerations. 
In particular, we evaluated the correlations between the model covariates, since high correlations can lead to a non-invertible design matrix resulting in invalid parameter estimates. This problem is known in statistics as \textit{multicollinearity} and more discussion can be found in \citet{belsley1991}.
In the analysis of real data presented in this work, we never observed a correlation coefficient exceeding $0.66$ between the asymmetry and width parameters and therefore the problem of multicollinearity appeared to be mitigated. 
We also investigated the statistical significant of the interactions between $A$ and the width of the CCF, and $A$ and the asymmetry of the CCF; those two interactions were not statistically significant.

%-----------------------------------------------------------------------------------------------------------------------------------------------
\section{Simulation Study} \label{sec:soap}
In order to evaluate the performance of the proposed SN approach for modelling the CCF and the benefit of using the proposed correction for stellar activity (see Eq.~\eqref{eq:RV:correction}), we begin by considering a simulation study using spectra generated from the Spot Oscillation And Planet 2.0 code \citep[SOAP 2.0,][]{Dumusque-2014b}.

For a given configuration of spots and faculae on the stellar surface, SOAP 2.0 outputs simulated CCFs as a function of rotational phase. 
The code also returns the RV and the FWHM estimated using a Normal model for the CCF, and the BIS SPAN obtained by calculating the bisector of the CCF. 
%\sout{SOAP 2.0 gives noiseless CCFs affected by stellar activity, which are used to compare the performances of SN and Normal models of the CCF.}

For the simulations discussed below, a star similar to the Sun was modelled with a solar disc of one solar radius seen equator-on, and a stellar rotational period set to 25.0 days.
The stellar effective temperature is set to 5778 K, and a quadratic limb-darkening relation with linear and quadratic coefficients 0.29 and 0.34, respectively \citep[][]{Oshagh-2013a, Claret-2011}. The temperature difference of the spot with the photosphere is $\Delta T_{spot} = -663$ K and the temperature difference of the facula depends on the centre-to-limb angle $\theta$,  $\Delta T_{facula} = 250.9 - 407.7\,\cos{\theta} + 190.9\,\cos^2 \theta$ K \citep[][]{Meunier-2010a}.
In order to make the result of the simulations more comparable to real data obtained with the HARPS spectrograph discussed in Sect.~\ref{sec:4}, the SOAP 2.0 CCFs were generated with a width of 40 \kms\, and considering initial spectra with resolution of Res =115'000. For the simulations presented in this section, we considered a S/N of 100.

\subsection{Facula} \label{sec:soap.faculae}

\begin{figure}[htbp]
\begin{center}
\includegraphics[width=3.6in]{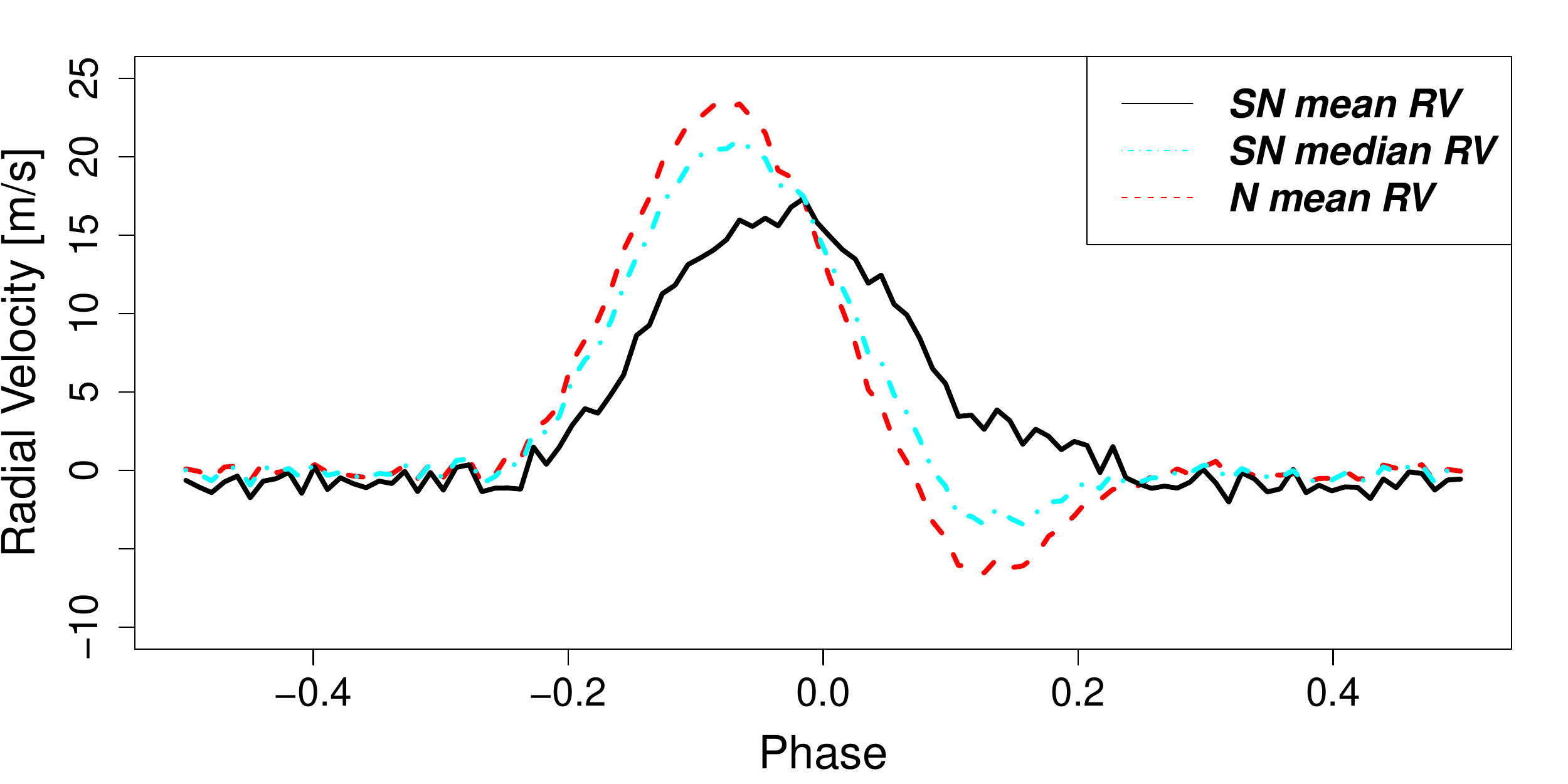} 
\caption{RV estimates for N mean RV (red dashed line),  SN mean RV (black line), and SN median RV (cyan dotted-dashed line). In this case, the CCFs were generated using SOAP 2.0 with an equatorial 3\% facula on the simulated Sun. The star does one full rotation between phase -0.5 and 0.5, with the facula being seen face-on for phase $0$. The variations observed in SN mean RV are notably different from the variations measured in SN median RV and N mean RV.  
}
   %\caption{(left) RVs changes as function of the orbital phase in the case in which a faculae is present on the photosphere of the star. SN mean RV seems to have the smallest spurious variations caused by the faculae. (right) Evaluation of the standard errors corresponding to the defined RVs. The standard errors retrieved for SN median RV are $10 \%$ smaller than the standard errors derived for RV. SN mean RV has the largest related uncertainties.}
    \label{fig:faculae}
\end{center}
\end{figure}

To see the impact of a facula on the parameters of different models of the CCF, we simulated the effect of an equatorial facula
%\jessi{Should a footnote be added here to explain that the SOAP 2.0 faculae are not simulated from actual faculae templates?} 
of size 3\% relative to the visible stellar hemisphere. The facula is face-on when the phase is 0. 
Note that a 3\% faculae is relatively large for the Sun; at maximum activity, large faculae generally have a size of 1\% \citep[e.g.][]{Borgniet-2015}. 
In Fig.~\ref{fig:faculae}, we compare the barycentric variation of the CCF as measured when fitting a Normal density and using its mean (N mean RV), and when fitting a SN density and taking its mean (SN mean RV) or median (SN median RV). We see that all the different estimates of the CCF barycenter present a signal of similar amplitude, however the signal obtained with SN mean RV is notably different from the two others with a maximum amplitude happening at a different phase.

\begin{figure*}[htbp]
\begin{center}
\includegraphics[height = 6in]{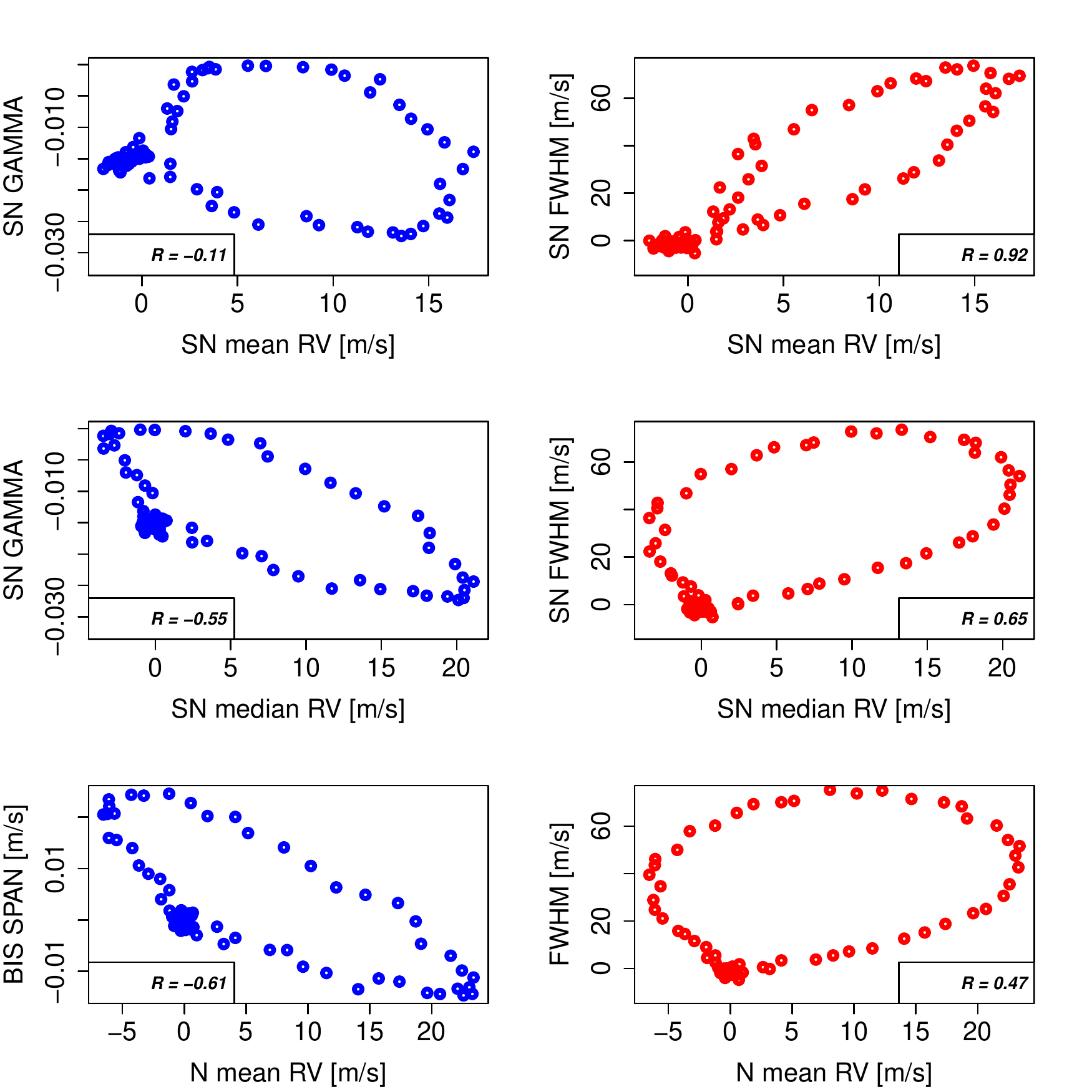} 
   \caption{(left) Correlations between the different asymmetry parameters and their corresponding RV estimates in the case of an equatorial 3\% facula on the simulated Sun. (right) Correlations between the different width parameters and their corresponding RV estimates for the same facula.
   In the presence of a facula, both the shape and the width of the CCF change as the star rotates, producing statistically significant correlations for all the cases except for the correlation between SN mean RV and SN GAMMA (P-value = 0.27).
   }
    \label{fig:faculae.corr}
\end{center}
\end{figure*}

Correlations between the different RV estimates and the different CCF asymmetry or width estimates are displayed in Fig.~\ref{fig:faculae.corr}. The correlation between $\gamma$ and SN mean RV, and $\gamma$ and SN median RV are weaker than the correlation between BIS SPAN and RV, with Pearson correlation coefficient values of $R$=-0.11, -0.55 and -0.61, respectively. Therefore, it seems that when looking at the CCF asymmetry, fitting a SN density to the CCF does not help. However, when looking at the correlations between the width of the CCF and the RV, the stronger correlation can be found between SN FWHM and SN mean RV ($R=0.92$). Then follows the correlation between SN FWHM and SN median RV ($R=0.65$), and finally FWHM and N mean RV ($R=0.47$). Over all the different correlations in the case of a facule, the strongest one is found between SN FWHM and SN mean RV, which demonstrates that fitting a SN density to the CCF can be helpful to better probe stellar activity.
%%
%{ {There is a stronger correlation between SN FWHM and SN mean RV ($R=0.92$) than between SN FWHM and SN median RV ($R=0.65$), but the correlation between FWHM and N mean RV is the weakest ($R=0.47$).}} 
%This first analysis shows that in the case of a facula, using some parameters from the SN can lead to stronger correlations than the ones obtained by the usual Normal parameters. Therefore, the SN parameters may better probe stellar activity. 
%We investigate this feature further in the following two sections, where we consider simulated data with a single spot and a spot plus a planet, and in Sec.~\ref{sec:4} with real observations.

\begin{figure*}[htbp]
\begin{center}
\includegraphics[height = 6in]{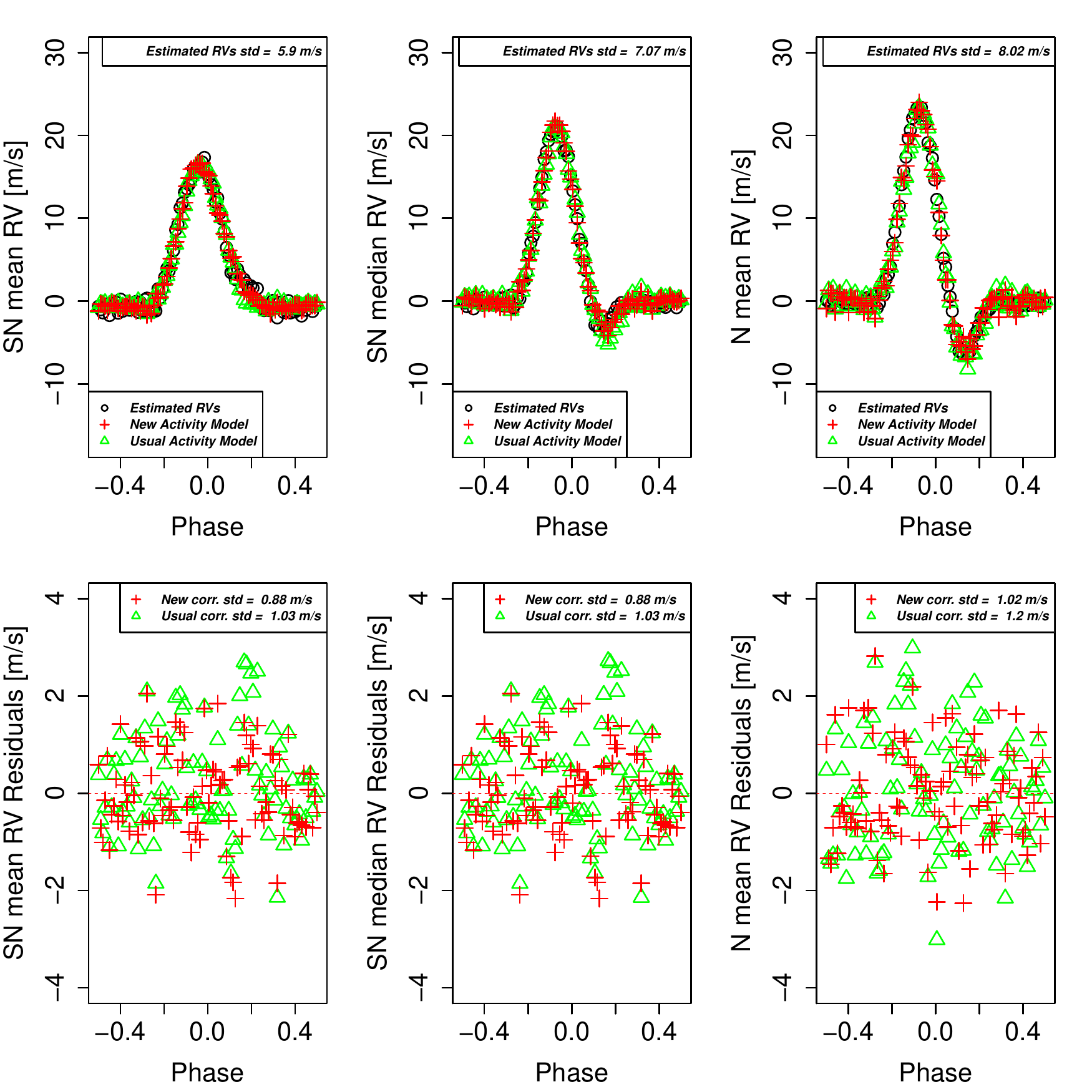} 
   \caption{(top) The spurious estimated RVs (black dots) caused by a facula in the simulated data using a Normal and a SN model, the estimated RVs using Eq.~\eqref{eq:RV:correction} (red pluses), and the estimated RVs using the usual correction for stellar activity (green triangles), based on $RV_{\text{activity}}=\beta_0+\beta_1 \gamma + \beta_2 \text{SN FWHM}$ for the SN fit and on $RV_{\text{activity}}=\beta_0+\beta_1 \text{BIS SPAN} + \beta_2 \text{FWHM}$ for the Normal fit.    
 (bottom) The residuals from the model fit using Eq.~\eqref{eq:RV:correction} (red pluses) and the residuals from the usual correction (green triangles). The standard deviations are also reported in the legend, and the residuals have a smaller systematic component when using the proposed model of Eq.~\eqref{eq:RV:correction} compared to the usual model.
The tests of statistical significance on the parameters are presented in Table \ref{table:faculae.test}.
}\label{fig:faculae.correction}
\end{center}
\end{figure*}

The RV variations displayed in Fig.~\ref{fig:faculae} are caused only by stellar activity, in this case a facula. We applied the activity correction proposed in Eq.~\eqref{eq:RV:correction} to evaluate the performance of the model in this setting.
The results of this correction are displayed in Fig.~\ref{fig:faculae.correction} and the statistical tests on the coefficients involved in Eq.~\eqref{eq:RV:correction} are summarized in Table \ref{table:faculae.test}. 
The proposed correction for stellar activity is able to account for the majority of the activity signal created by a facula, with a $R^2$ larger than $0.98$.
%In addition, the rms of the different estimates of the RV reduces from about { {6 \ms}} before correcting for stellar activity to values below { {1.2 \ms}} after correcting for stellar activity.
%We see a slightly smaller rms after correction when using the SN parameters compared to the Normal parameters.
In addition, the proposed linear model allows to reduce the activity effect from a RV rms of 8.02 \ms to 1.02 \ms when considering the CCF contrast and FWHM, and the BIS SPAN. When using the parameters derived from the SN, the improvements for SN median RV and SN mean RV are 7.07 \ms down to 0.88 \ms and 5.9 \ms down to 0.88 \ms, respectively.
When comparing the correction proposed in Eq.~\eqref{eq:RV:correction} with what is generally used (i.e. a linear combination of only the asymmetry and width parameters), we see that the proposed correction is able to reduce the rms of the RV residuals by an additional 14.5 - 15\%. Looking at the significance of the coefficients in Table~\ref{table:faculae.test}, we observe that all the parameters corresponding to the SN variables are statistically significant. When using the Normal variables, the parameters $\beta_{3}$ and $\beta_{4}$ are not statistically helpful for the correction.

\begin{table}
\begin{center}
\caption{P-values for the estimated coefficients from the model in Eq.~\eqref{eq:RV:correction} for correcting stellar activity induced by an equatorial 3\% facula on the simulated Sun. All the parameters corresponding to the SN variables are statistically significant. When using the Normal variables, the parameters $\beta_{3}$ and $\beta_{4}$ are not statistically helpful for the correction. The estimated $R^{2}$ show that the proposed correction for stellar activity explains the vast majority of the spurious variability present in the different RV estimates.}
\label{table:faculae.test}
\begin{tabular}{|c|c|c|c|}
\hline
Parameter          & N mean RV         &   SN mean RV &   SN median RV \\
\hline
$\beta_{0}$            &    $0.3e-4$    & $0.081$ & $0.34e-3$ \\
\hline
$\beta_{1}$            &    $2.01e-8 $    & $5.11e-5 $ & $4.74e-5 $ \\
\hline
$\beta_{2}$            &     $2.22e-16$   &  $1.66e-5 $ & $2.22e-16 $\\
\hline
$\beta_{3}$            &     $0.48$   &  $0.013$ & $0.012$\\
\hline
$\beta_{4}$            &     $0.28$   &  $0.31e-3$ & $0.22e-3$\\
\hline
$R^{2}$      &     $0.9839$    &  $0.9801$ & $0.9844$  \\
\hline
\end{tabular}
\end{center}
\end{table}

\subsection{Spot} \label{sec:soap.spot}

\begin{figure}[htbp]
\begin{center}
\includegraphics[width=3.6in]{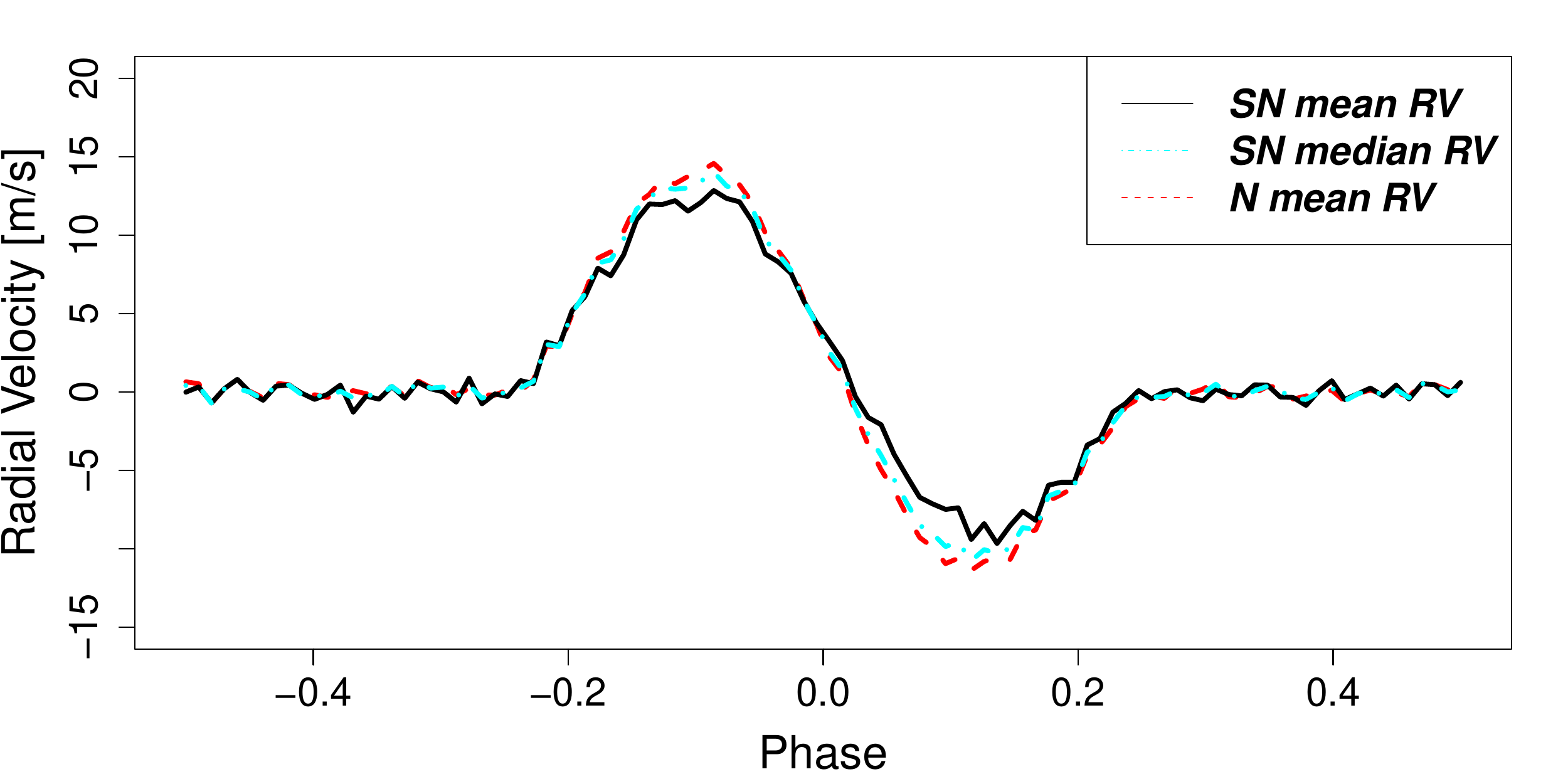} 
\caption{RV estimates for N mean RV (red dashed line),  SN mean RV (black line) or SN median RV (cyan dotted-dashed line) using CCFs generated from SOAP 2.0 with an equatorial 1\% spot on the simulated Sun. The star does one full rotation between phase -0.5 and 0.5, with the spot being seen face-on at phase 0. The SN mean RV seems to have the smallest spurious variations caused by the spot.}
   %\caption{(left) RVs changes as function of the orbital phase in the case in which a spot is present on the photosphere of the star. SN mean RV seems to have the smallest spurious variations caused by the faculae. (right) Evaluation of the standard errors corresponding to the defined RVs. The standard errors retrieved for SN median RV are $10 \%$ smaller than the standard errors derived for RV. SN mean RV has the largest related uncertainties.}
    \label{fig:spot}
\end{center}
\end{figure}

Next we consider the effects on the CCF model parameters due to the presence of an equatorial spot of size 1\% relative to the visible stellar hemisphere. 
The spot is face-on when the phase is 0. 
Note that this is a large spot for the Sun, as large spots are generally on the order of 0.1\% \citep[e.g.][]{Borgniet-2015}. 
In Fig.~\ref{fig:spot}, we show the barycentric variations of the CCF induced by this simulated spot. 
In contrast to the case of the facula, all the different estimates of the CCF barycenter for the spot have the same shape in variation; however, the amplitude for SN mean RV is slightly smaller.

\begin{figure*}[htbp]
\begin{center}
\includegraphics[height = 6in]{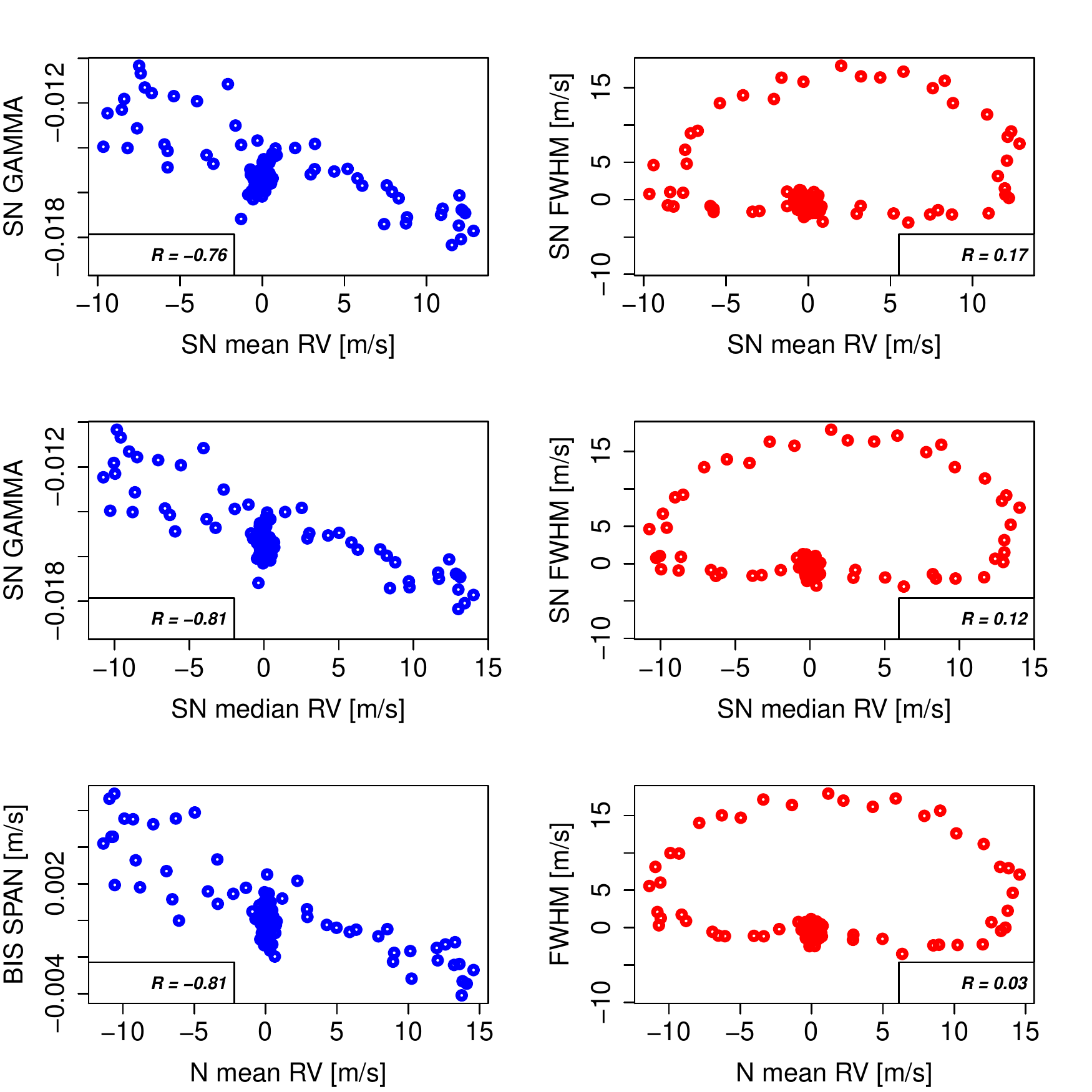} 
   \caption{(left) Correlations between the different asymmetry parameters and their corresponding RV estimates in the case of an equatorial 1\% spot on the simulated Sun. (right) Correlations between the different width parameters and their corresponding RV estimates for the same spot. In the presence of a spot, both the shape and the width of the CCF change as the star rotates. However, only the asymmetry produces a statistically significant correlation with the different RV estimates. The width parameters and their corresponding RV estimates present weak correlations and, in general, much weaker correlations compared to the results obtained when an equatorial 3\% facula is present on the simulated Sun.}
    \label{fig:spot.corr}
\end{center}
\end{figure*}

Fig.~\ref{fig:spot.corr} shows the correlations between the asymmetry parameters and the different estimates of the CCF barycenter (i.e. SN mean RV, SN median RV and N mean RV). The correlation between $\gamma$ and SN median RV and the correlation between the BIS SPAN and N mean RV are the strongest ($R=-0.81$) followed by the $\gamma$ - SN mean RV correlation ($R=-0.76$).
The correlations between the width and the CCF barycenter draw circles and no significant correlation is observed. Unlike the facula scenario, when considering a spot simulated from SOAP 2.0 and a S/N of 100 the SN parameters do not appear to better probe stellar activity than the Normal parameters.
%\sout{Similar to the case with a facula, some parameters of the SN gives stronger correlations compared to the Normal parameters.}}

\begin{figure*}[htbp]
\begin{center}
\includegraphics[height = 6in]{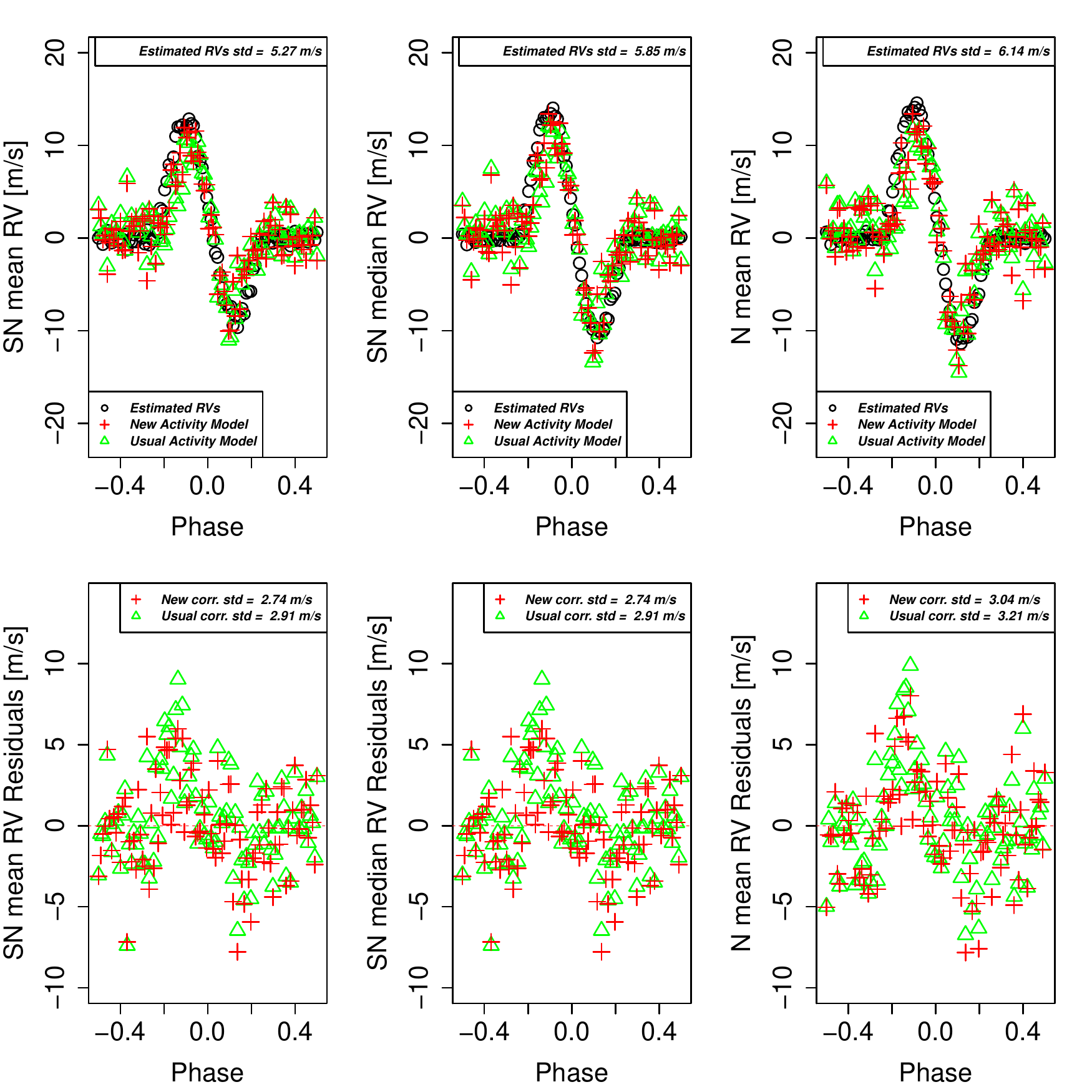}
   \caption{(top) The spurious estimated RVs (black dots) caused by a spot in the simulated data, the estimated RVs using Eq.~\eqref{eq:RV:correction} (red pluses) and the estimated RVs using the usual correction for stellar activity (green triangles), based on $RV_{\text{activity}}=\beta_0+\beta_1 \gamma + \beta_2 \text{SN FWHM}$ for the SN fit and on $RV_{\text{activity}}=\beta_0+\beta_1 \text{BIS SPAN} + \beta_2 \text{FWHM}$ for the normal fit.
 (bottom) The residuals from the model fit using Eq.~\eqref{eq:RV:correction} (red pluses) and the residuals from the usual correction (green triangles). 
 The standard deviations are also reported in the legend, and the residuals have a smaller systematic component when using the proposed model compared to the usual model.
The tests of statistical significance on the parameters are presented in Table~\ref{table:spot.test}.
}
    \label{fig:spot.correction}
\end{center}
\end{figure*}

As before, the original RV estimates are corrected using Eq.~\eqref{eq:RV:correction}. 
The results of this correction are displayed in Fig.~\ref{fig:spot.correction} and the statistical tests on the coefficients involved in Eq.~\eqref{eq:RV:correction} are summarized in Table~\ref{table:spot.test}. 
In the case of a spot, the proposed correction is not able to perform as well as for the facula, and $R^2$ values for the linear combination are between 0.7 and 0.8. 
The correction is able to mitigate stellar activity from a N mean RV rms of 6.14\ms down to 3.04\ms. The corresponding values for SN median RV and SN mean RV are 5.85\ms down to 2.74 \ms and 5.27\ms down to 2.74\ms, respectively.
When comparing the activity correction proposed in this paper with what is commonly used, which means only a linear combination of the width and asymmetry of the CCF, we see that our solution is capable of reducing the RV residual rms by 5.3 - 5.8 \%.
The Normal or SN parameters involved in Eq.~\eqref{eq:RV:correction} are statistically significant to explain the activity signal as seen in Table~\ref{table:spot.test}, except the width of the CCF $\beta_3$ and the interaction term $\beta_4$.
%This is not surprising when looking at the circle shape drawn when plotting the width as a function of the RV in Fig.~\ref{fig:spot.corr}. 

\begin{table}
\begin{center}
\caption{P--values for the different coefficients used in Eq.~\eqref{eq:RV:correction} for the correction of stellar activity induced by an equatorial 1\% spot on the simulated Sun. All the parameters corresponding to the Normal or SN parameters are statistically significant to explain the spurious RV variations caused by this spot, except for the width of the CCF and the variable that evaluates the interaction between the width and the asymmetry of the CCF. The proposed correction is not able to perform as well as for the facula, and $R^2$ values for the linear combination are smaller than 0.8.
} 
\label{table:spot.test}
\begin{tabular}{|c|c|c|c|}
\hline
Parameter          & N mean RV         &   SN mean RV &   SN median RV \\
\hline
$\beta_{0}$            &    $0.0049$    & $0.012$ & $0.014$ \\
\hline
$\beta_{1}$            &    $0.0021$    & $0.0011$ & $0.0011$ \\
\hline
$\beta_{2}$            &     $2.22e-16$   &  $2.22e-16$ & $2.22e-16$\\
\hline
$\beta_{3}$            &     $0.92$   &  $0.76$ & $0.76$\\
\hline
$\beta_{4}$            &     $0.34$   &  $0.41$ & $0.41$\\
\hline
$R^{2}$      &     $0.7553$    &  $0.7283$ & $0.7799$  \\
\hline
\end{tabular}
\end{center}
\end{table}

\subsection{Spot and planet} \label{sec:soap.spot.planet}

The final simulation includes a planetary signal influencing the CCF along with the 1\% spot modelled previously (see Sec.~\ref{sec:soap.spot}). 
The purpose of this example is to check if we are able to disentangle these two different sources of variations when using the parameters derived using a Normal or a SN model for the CCF. In this scenario the planet is injected with a semi-amplitude of 10 \ms with no eccentricity and with a period corresponding to one-third of the stellar rotational period, i.e. one-third of 25 days.

\begin{figure}[htbp]
\begin{center}
\includegraphics[width = 3.6in]{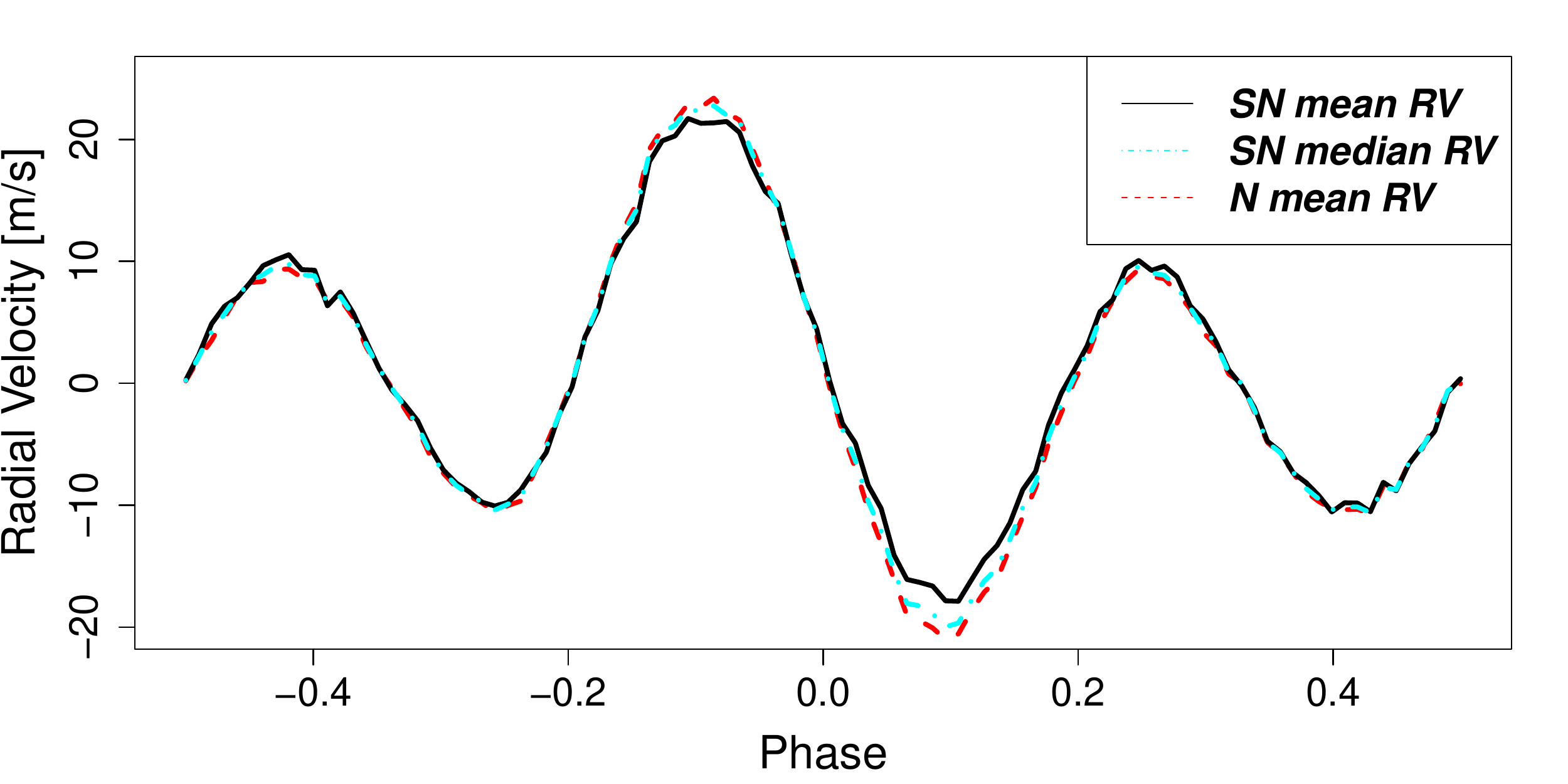} 
 \caption{RV estimates for N mean RV (red dashed line),  SN mean RV (black line) or SN median RV (cyan dotted-dashed line). In this case, the CCFs have been generated using SOAP 2.0, considering an equatorial 1\% spot on the simulated Sun in addition to a planet with a period of one-third of the rotational period of the star and with an amplitude of 10\,\ms. The star does one full rotation between phase -0.5 and 0.5, with the spot being seen face-on at phase 0.}
   %\caption{(left) RVs changes as function of the orbital phase in the case in which a spot is present on the photosphere of the star and a planet is injected. N mean RV seems to have the largest variations caused by the combined action of spot and planet. (right) Evaluation of the standard errors corresponding to the defined RVs. The standard errors retrieved for SN median RV are $10 \%$ smaller than the standard errors derived for RV. SN mean RV has the largest related uncertainties.}
    \label{fig:spot.plus.planet}
\end{center}
\end{figure}

Fig.~\ref{fig:spot.plus.planet} shows the variations observed in the CCF barycenter parameters. As in the case of the spot, all RV estimates show similar variations, with SN mean RV showing a slightly smaller amplitude.

\begin{figure*}[htbp]
\begin{center}
\includegraphics[height = 6in]{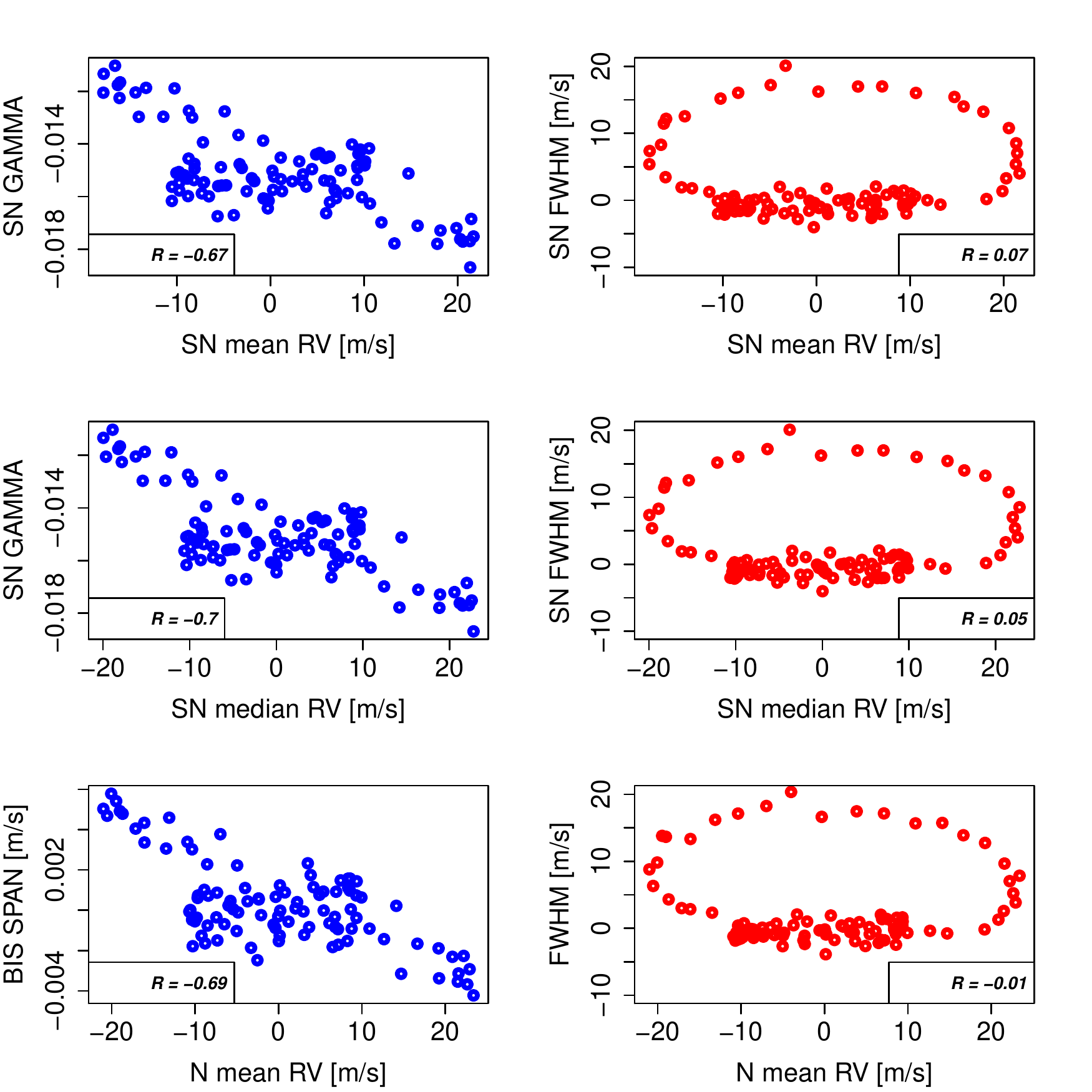} 
   \caption{Evaluation of the correlation between the RVs and the asymmetry parameters of the simulated data with a 1\% spot and an injected planetary signal.  The shape of the CCF changes as the spot moves, producing statistically significant correlations only between the estimated RVs and the asymmetry parameter. The correlations between the estimated RVs and the width parameter of the CCF are weaker than the case with only a spot.}
    \label{fig:spot.plus.planet.corr}
\end{center}
\end{figure*}

The correlation between the different CCF parameters are displayed in Fig.~\ref{fig:spot.plus.planet.corr}. 
The correlations are weaker than in the case of the spot due to the planet inducing changes in RV without affecting the width or the asymmetry of the CCF.  However, the order of the strength of the correlations between the CCF asymmetry parameters and RV are comparable with the ones obtained for the spot-only model: $\gamma$--SN median RV has the strongest correlation ($R=-0.7$), followed by the correlation between BIS SPAN--N mean RV ($R=-0.69$) and finally by the correlation between $\gamma$--SN mean RV ($R=-0.67$). 
The patterns observed in the width-RV phase space correlations in Fig.~\ref{fig:spot.plus.planet.corr} follow a circle, similar to the spot-only model; no substantial correlation is observed between those two parameters.

In order to correct the estimated RVs from the spurious variation caused by the spot, the proposed model for correcting the activity is added to a planetary signal model that takes into account the RV variations caused by a planet. The observed RVs can therefore be modelled as a combination of the activity and the planetary signals:
\begin{equation}
RV= RV_{\text{activity}} + RV_{\text{planet}},
\label{eq:RV:correction.overall}
\end{equation}
where $RV_{\text{activity}}$ can be found in Eq.~\eqref{eq:RV:correction}, and $RV_{\text{planet}}$, in the case with no eccentricity, can be modelled by the following sinusoidal function:
\begin{equation}
RV_{\text{exoplanet}}= K \sin \left(\frac{2 \pi}{P} (t - t_{0})\right),
\label{eq:RV:correction.planet}
\end{equation}
with amplitude $K$, orbital period $P$, and an epoch at the periapsis $t_{0}$.  The previous three unknown parameters define the planetary orbit.

%We note that the p--value associated with the amplitude parameter $K$ is particularly relevant for rejecting or not rejecting the assumption about the presence of an orbiting companion. Moreover we note also that Eq.~\eqref{eq:RV:correction.planet} is highly non linear, meaning that the estimation of all the parameters involved in Eq.~\eqref{eq:RV:correction.overall} has to be done numerically. %We used non linear least squares and the results are displayed in Fig.~\ref{fig:spotplanet.correction}. We can see how in this case we are able to disentangle the spurious variations in RVs caused by stellar activity from the pure Doppler-shift due to the planet. 
The proposed model from Eq.~\eqref{eq:RV:correction.overall} was fitted to the RV data and the results of the estimated model are summarized in Table~\ref{table:spotplanet.test}.   Except for the intercept $\beta_0$, the width of the CCF $\beta_3$, and the interaction term $\beta_4$ that evaluates the interaction between the width and the asymmetry of the CCF, all the other Normal or SN parameters are significantly useful to explain the RV variation induced by a spot plus a planet. We also observe that the RV residuals, once corrected for stellar activity and the presence of the planet, are smaller in terms of rms when using the SN variables (rms = 2.29 \ms) rather than the Normal ones (rms = 2.80 \ms).
%\begin{figure*}[htbp]
%   \centering
%\includegraphics[height = 4in]{SPOT_PLANET_NEW_CORRECTION.pdf} 
%   \caption{Set of  variations in RVs estimated using a Normal and a SN fit before and once corrected from stellar activity. In this case there are spurious variations caused by the spot and pure Doppler-shift due to the planet. The correction is done using Eq.~\eqref{eq:RV:correction.overall} and the estimated parameters are presented in Table \ref{table:spotplanet.test}. In this case, by solving Eq.~\eqref{eq:RV:correction.overall}, we are able to completely disentangle the spurious variations in RVs caused by the presence of the spot from the pure dopplershift caused by the exoplanet.}
%    \label{fig:spotplanet.correction}
%\end{figure*}

\begin{table}
\begin{center}
\caption{P-values for the different coefficients used in Eq.~\eqref{eq:RV:correction} for the correction of stellar activity induced by an equatorial 1\% spot on the simulated Sun, and a planet with a period of one-third the rotational period of the star and a semi-amplitude  of 10\,\ms. All the parameters corresponding to the Normal or SN variables are statistically significant, except the intercept $\beta_0$, the width of the CCF $\beta_3$ and the interaction term $\beta_4$ that evaluates the interaction between the width and the asymmetry of the CCF. Once corrected for stellar activity and the presence of the planet, the residuals are smaller in terms of rms when using the SN variables. Note that because we used nonlinear least squares to fit the best model, the residual rms rather than the $R^2$ is displayed as a reference.}
\label{table:spotplanet.test}
\begin{tabular}{|c|c|c|c|}
\hline
Parameter          & N mean RV         &   SN mean RV &   SN median RV \\
\hline
$\beta_{0}$            &    $0.95$    & $0.076$  & $0.21$ \\
\hline
$\beta_{1}$            &    $0.0020$    & $0.11e-4$  & $0.12e-4$ \\
\hline
$\beta_{2}$            &     $2.22e-16$   & $2.22e-16$ & $2.22e-16$\\
\hline
$\beta_{3}$            &     $0.92$   &  $0.60$  & $0.61$\\
\hline
$\beta_{4}$            &     $0.34$   &  $0.27$ & $0.27$\\
\hline
K            &     $2.22e-16$   &  $2.22e-16$   & $2.22e-16$ \\
\hline
P            &     $2.22e-16$   &  $2.22e-16$ & $2.22e-16$ \\
\hline
$t_{0}$            &     $2.22e-16$   &  $2.22e-16$ & $2.22e-16$ \\
\hline
$\text{Residuals}$      &     $2.80 \ms$    &  $ 2.29 \ms$ & $2.29 \ms$  \\
\hline
\end{tabular}
\end{center}
\end{table}

\section{Real data application} \label{sec:4}

The analysis of the simulated SOAP 2.0 data in the previous section were helpful in assessing the performance of the proposed methodology in a setting where the ground truth was known. We found with the simulated data that the parameters derived by the SN had correlations comparable to those derived by the Normal; in the case of a facula, the SN parameters had higher correlations for the FWHM than those of the Normal. In this section we present an analysis conducted on real observations, in particular, the star Alpha Centauri B, and compare the performance when fitting a CCF using the SN density defined in Sec. \ref{sec:3} with the commonly employed approach based on fitting a Normal density for estimating the RV and retrieving the BIS SPAN for evaluating the asymmetry of the CCF. 
%to retrieve the RV and width of the CCF and calculating the bisector to derive the asymmetry parameter BIS SPAN. 
Four other stars have been analyzed with the proposed method and details can be found in Appendix \ref{appendix}. 
For all the stars considered in the presented work, only CCFs that were derived from spectra that had at least a S/N of 10 at 550 nm were selected. 

% A comparison with the results obtained by the classic approach is done, where the RV is estimated by retrieving the mean of the Normal density used to fit the CCF, along with the FWHM of the Normal density and the BIS SPAN or the other asymmetric parameters defined in \citet{Figueira-2013}. The latter parameters are calculated separately from the Normal fit that leads to the set of RVs of the star.

\subsection{Comparison for Alpha Centauri B of the different CCF parameters derived with the Normal and the Skew Normal} \label{sec:alphacentb}

A total of $1808$ CCFs that were derived from the spectra of Alpha Centauri B taken in 2010 by the HARPS spectrograph have been analyzed. Note that more observations were carried out that year, however only the data that were not significantly affected by contamination from Alpha Centauri A were used \citep[see][]{Dumusque-2012}. 
The selected observations represent arguably, among all RV data existing, the best sampled and most precise RV data set showing strong solar-like activity signal \citep{Dumusque:2018aa, Thompson-2017}.

\begin{figure*}[htbp]
\begin{center}
\includegraphics[height = 3.6in]{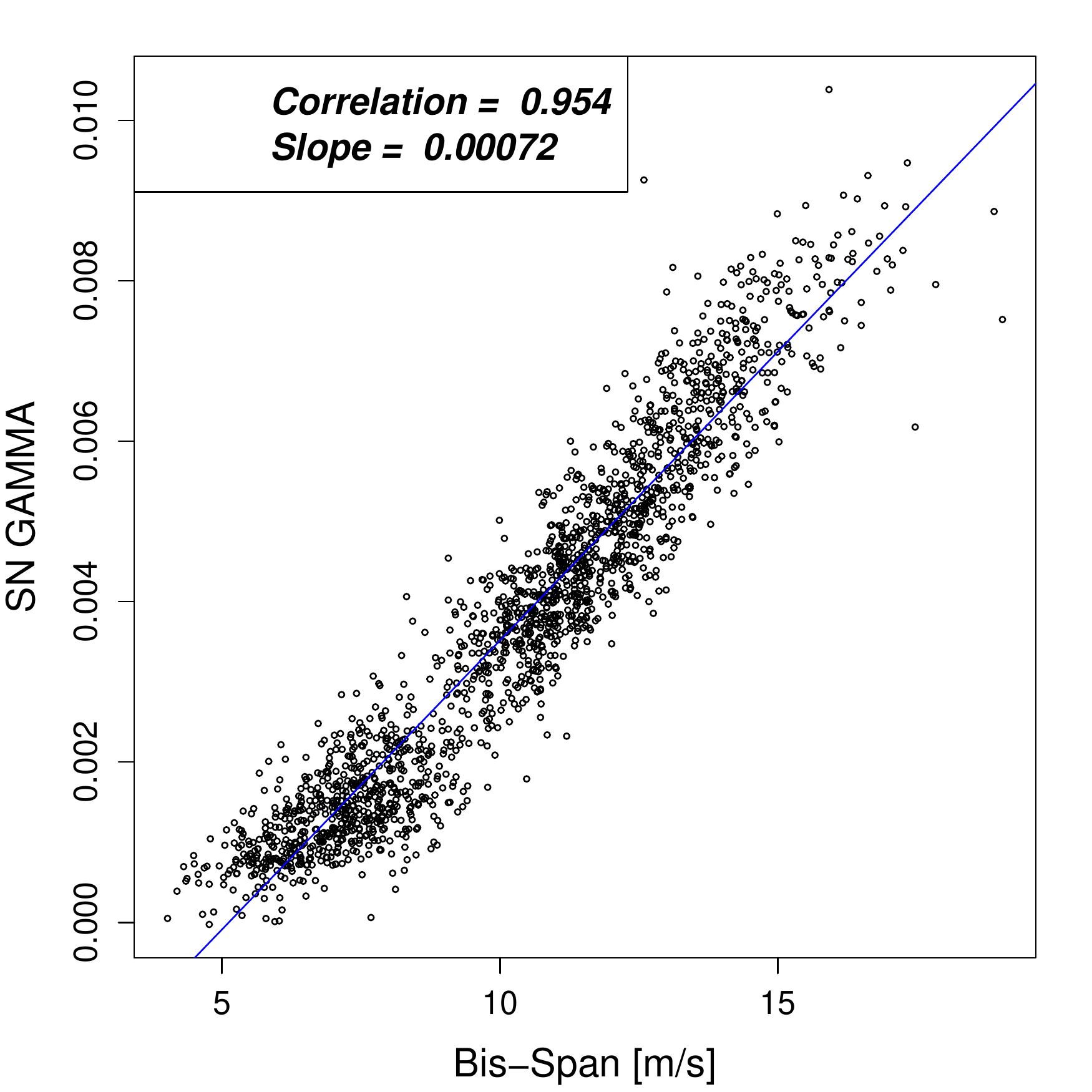} 
\includegraphics[height = 3.6in]{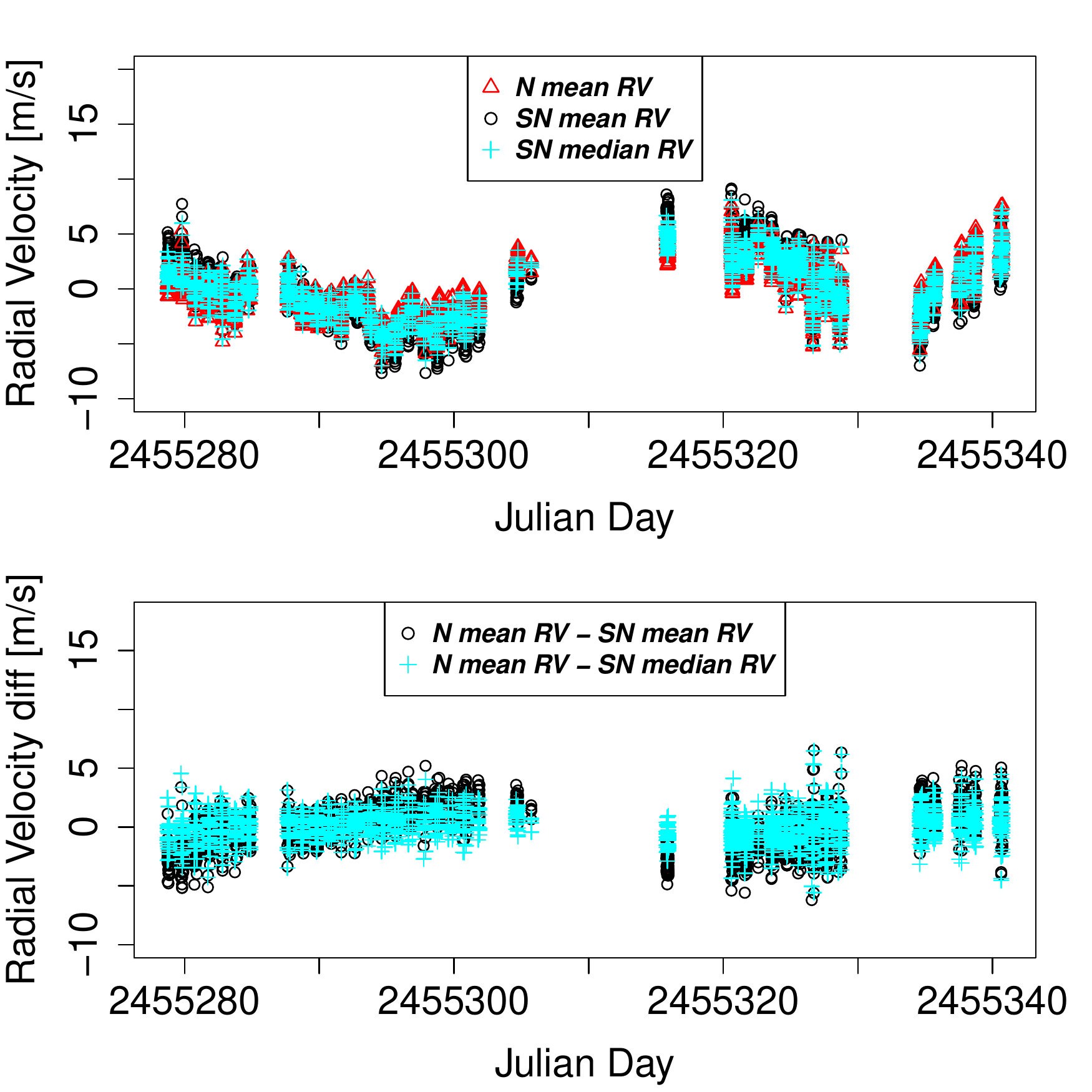} 
   \caption{(left) Correlation between $\gamma$ and the BIS SPAN for Alpha Centauri B. The strong correlation suggests these two parameters are similarly measuring the asymmetry. (top right) RVs as function of Julian Day for Alpha Centauri B in 2010. The RVs are estimated using the mean of a Normal fitted to the CCF (red triangles), or the mean (black circles) or median (cyan pluses) of a SN density fitted to the CCF. (bottom right) Differences between the RVs estimated with the Normal density and those from the SN density.}
   \label{fig:alphacent:corr.gamma}
\end{center}
\end{figure*}

First, the correlation between $\gamma$ and BIS SPAN is evaluated.
In the left panel of Fig.~\ref{fig:alphacent:corr.gamma}, we see that the relationship between $\gamma$ and the BIS SPAN is linear, with a slope equal to $0.00072$ and a strong Pearson correlation coefficient of $R=0.954$. This strong correlation suggests that both $\gamma$ and BIS SPAN are measuring similar asymmetries for the CCFs. 
It also provides a conversion between the dimensionless $\gamma$ parameter into \ms\, using the slope of $0.00072$ \ms.
%\xavier{As $\gamma$ has no units, this will allow us to compare the amplitude of the activity signal seen in $\gamma$ and in BIS SPAN, by using the slope of this correlation as a scaling factor}.

%

The right plot of Fig.~\ref{fig:alphacent:corr.gamma} displays the comparison between the RVs estimated using the SN density and the Normal density. 
The amplitude of the activity signal is slightly stronger for SN mean RV (in the top-right plot the black circles of SN mean RV tend to show more variability), while the signal measured using N mean RV or SN median RV are comparable. 
%
%\begin{figure*}[htbp]
%   \centering
%\includegraphics[height = 3in]{HD12862_[2]RadialVelocityDifferences.pdf} 
%   \caption{(top) RVs as function of Julian Day for Alpha Centauri B. The RVs are retrieved using the mean of the Normal (red triangles), SN mean RV (black circles), SN median RV (cyan pluses). (bottom) RV differences between Normal RV and SN mean RV (black circles) and between Normal RV and a SN median RV (cyan pluses).}
%   \label{fig:alphacent:diff:RV}
%\end{figure*}
%

\begin{figure*}[htbp]
\begin{center}
\includegraphics[height = 6in]{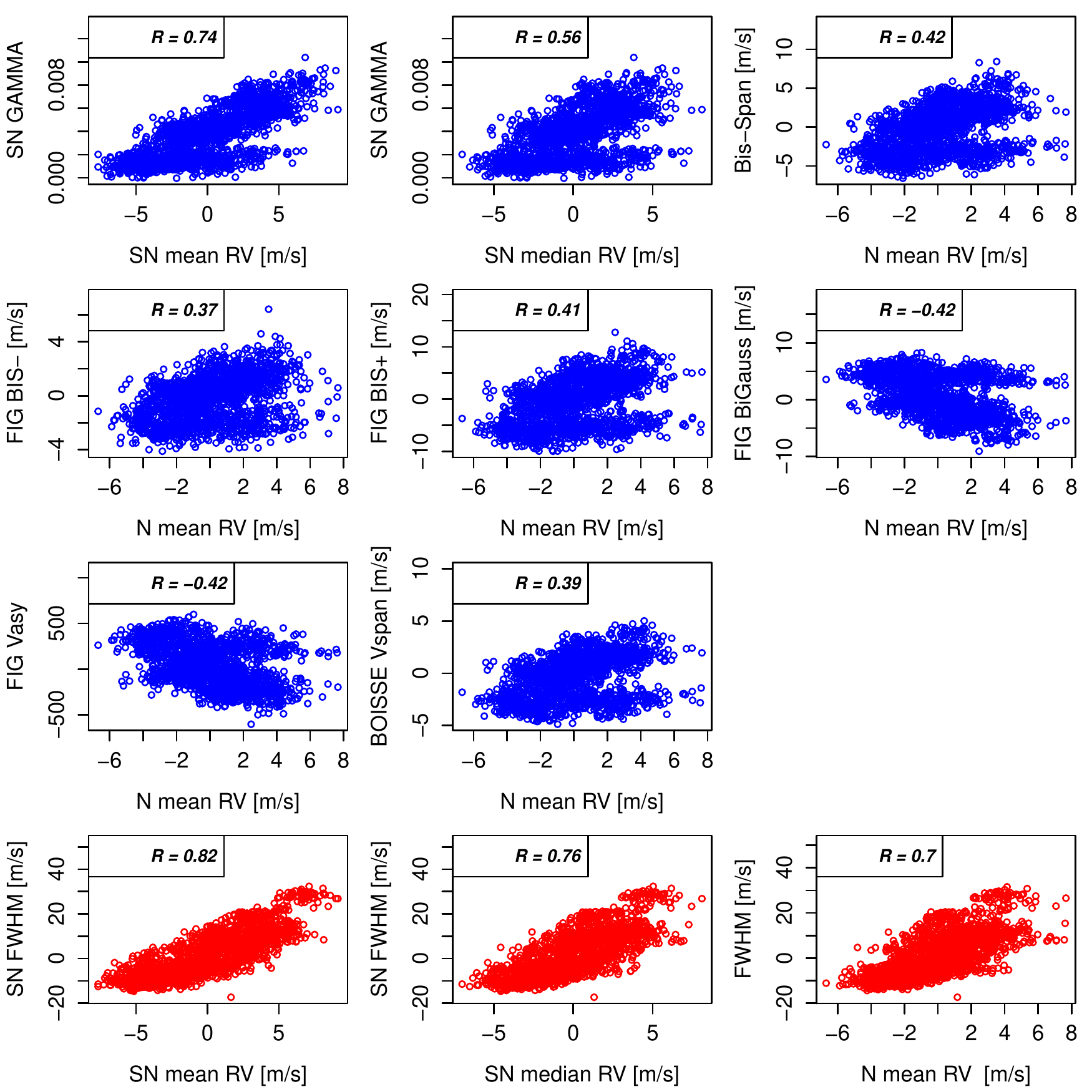}  
   \caption{(top three rows) Correlations between the asymmetry parameters and their corresponding estimated RVs for Alpha Centauri B. 
(bottom row) The correlation between the FWHM and the estimated RVs. The correlations are stronger when using parameters derived from the SN fit than the Normal one. The estimated $R$'s are all statistically significant.} 
   \label{fig:alphacent:corrPlot}
\end{center}
\end{figure*}

Similar to the analyses presented in Sec.~\ref{sec:soap}, in Fig.~\ref{fig:alphacent:corrPlot} we compare the correlations between the asymmetry or the width parameters of the CCF and the RV. 
For this analysis, we also include the asymmetry parameters derived in \citet{Boisse-2011}, $V_{span}$ and in \citet{Figueira-2013}, BIS-, BIS+, Bi Gauss and $V_{asy}$, as these authors found those asymmetry parameters more correlated with the RVs than BIS SPAN. It is clear in the case of Alpha Centauri B that the correlation found between $\gamma$ and SN mean RV is the strongest. 
The Pearson correlation coefficient is $R=0.74$, while the next strongest is $R=0.42$ for all the other asymmetry-N mean RV correlations.
The correlations between the width and the RV estimates for Alpha Centauri B is also the strongest for the SN parameters, with $R=0.82$ for SN FWHM-SN mean RV, compared to $R=0.70$ for FWHM-N mean RV.
%
%The correlations between $\gamma$ and SN mean RV, and also SN FWHM and SN mean RV
%
%Because the median is a more robust index than the mean, the correlation between $\gamma$ and SN median RV is not as large as the correlation between $\gamma$ and SN mean RV, but it is nonetheless $1.5$ times larger than the correlation between the other common asymmetry parameters and their corresponding RVs. In other words, changes in the asymmetry of the CCF are better captured when using the SN mean RV. The correlation between FWHM and the RVs, either by using SN mean RV or SN median RV, is as well stronger when fitting a SN density rather than a Normal. All the correlations are statistically different from $0$. Recalling the analyses presented in Sec. \ref{sec:soap}, we could infer that Alpha Centauri B is dominated by faculae, because the correlations between the RVs and the width of the CCF are strong (in particular the correlation between SN mean RV and SN FWHM is $0.817$).
%%

{  When comparing the correlation between the different Normal and SN parameters in the case of the real data of Alpha Centauri B (see Fig.~\ref{fig:alphacent:corrPlot}) with the correlations obtained in the SOAP 2.0 simulations (see Sec.~\ref{sec:soap.spot.planet}), we observe some significant differences. The correlations between the different parametrisations of the CCF asymmetry and barycentre do not match between the real and simulated data. In the real case, the correlations between $\gamma$ and SN mean RV, $\gamma$ and SN median RV and BIS SPAN and N mean RV are all positives. In the case of the SOAP 2.0 simulated data for a facula or a spot, we always find negatives correlations. It is therefore not possible to reproduce with SOAP 2.0 the CCF asymmetry variations observed in real observations. On the contrary, the correlations between the CCF width and barycentre match between the real data and a SOAP 2.0 simulated facula. It seems therefore that SOAP 2.0 simulation are able to correctly model the width variation of the CCF, however not its asymmetry. This is probably because in the SOAP 2.0 simulation, a facula is modelled using the same spectrum as a spot, with only a different flux. It is well known that facula have a different temperature than spots, and therefore a spectrum that significantly differs \citep[e.g.][]{Cavallini-1985a}. Looking at other stellar activity simulation like StarSim \citep[][]{Herrero:2016aa}, it seems that simulating a positive correlation between the CCF asymmetry and barycentre is not possible with current tools, and some progress are still to be made.}

\begin{figure*} 
\begin{center}
\includegraphics[height = 6in]{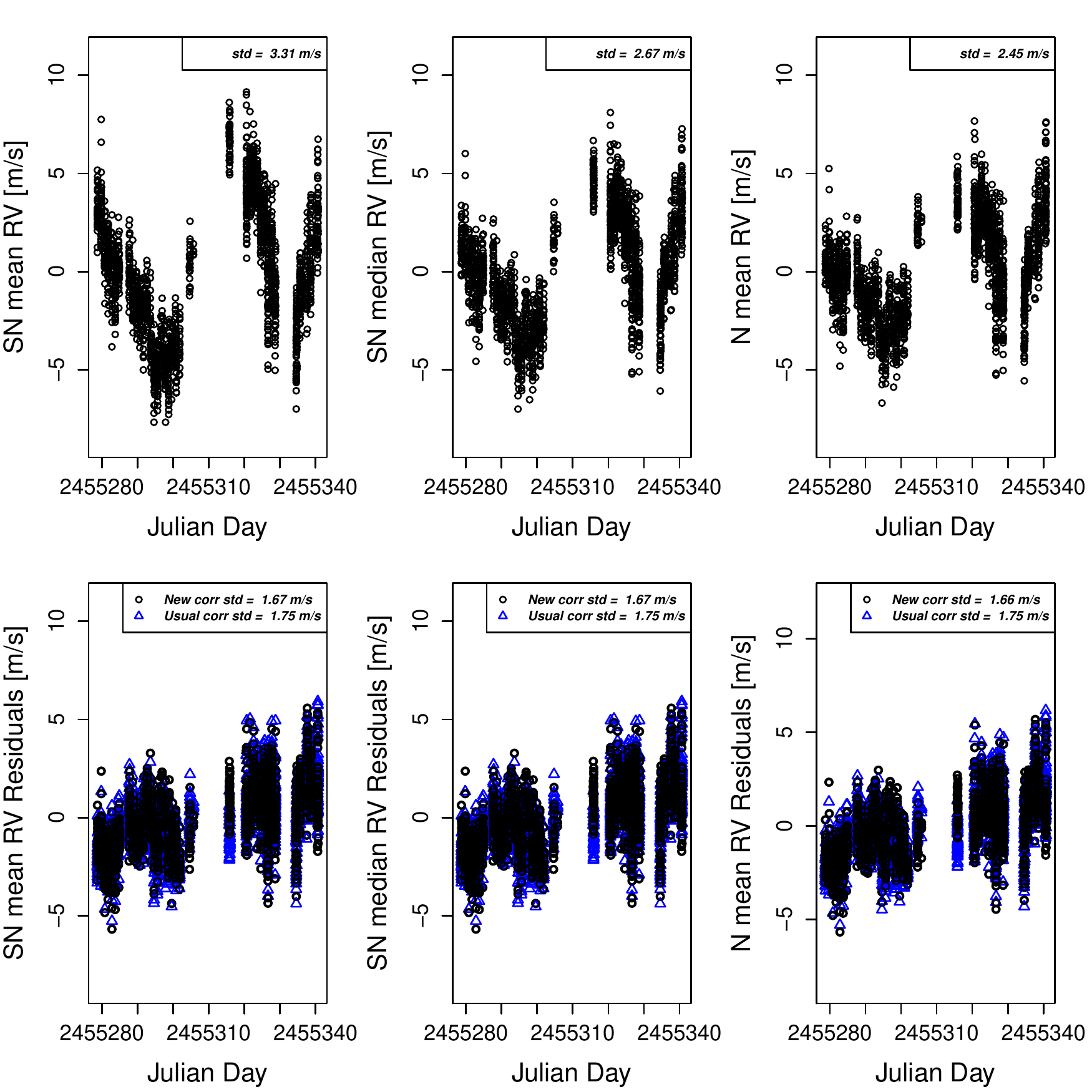} 
   \caption{(top) The RVs (black dots) for Alpha Centauri B estimated using a SN and a Normal fit.
 (bottom) The residuals from the model fit using Eq.~\eqref{eq:RV:correction} (New corr. std, black dots) and the residuals from the usual correction (Usual corr. std, blue triangles), based on $RV_{\text{activity}}=\beta_0+\beta_1 \gamma + \beta_2 \text{SN FWHM}$ for the SN fit and on $RV_{\text{activity}}=\beta_0+\beta_1 \text{BIS SPAN} + \beta_2 \text{FWHM}$ for the normal fit. The residuals have a smaller systematic component when using the proposed model of  Eq.~\eqref{eq:RV:correction} (black dots) compared to the usual model (blue triangles).}
\label{fig:alphacent:correctionRV}
\end{center}
\end{figure*}

Results illustrating the performance of the stellar activity correction proposed in Sec.~\ref{sec:31} are displayed in Fig~\ref{fig:alphacent:correctionRV}. 
For Alpha Centauri B, the RV estimated with SN mean RV has a rms that is 35\% larger than the rms of the RV estimated with the N mean RV, and the rms of SN median RV is 9\% larger than that of the N mean RV.
Even though we see these differences in the estimated RV, once we correct for stellar activity using Eq.~\eqref{eq:RV:correction}, 
the rms of the residuals are essentially the same for all three approaches.
%In the best case, for SN mean RV, we reduce the stellar activity signal by a factor of 2, while in the worst case, N mean RV, only by a factor 1.5. 
Although the correlations between the different parameters from the SN density are more sensitive to stellar activity than those obtained with a Normal density fit,
the proposed linear model that corrects for stellar activity does not necessarily perform better in the SN case than in the Normal case. 
The new correction for stellar activity proposed in Sec.~\ref{sec:31} performed only slightly better than the usual correction that uses only a linear combination of the width and the asymmetry of the CCF.

%%with the usual correction that uses only a linear combination of the width and the asymmetry of the CCF, we end up with very similar results. The new correction is however able to get RV rms that are 6\,\cms\,smaller than the usual correction.
%it seems that the parameters derived using the SN density are more sensitive to stellar activity than with the Normal density, after correction using the proposed model, the methods appear to be similarly successful at addressing the stellar activity signal.
%we are not able to correct better for stellar activity signal using a linear combination of the different CCF parameters.

The results of the statistical tests of the different parameters used for correcting activity can be found in Table~\ref{table:alphacent.test}.  The BIS SPAN (coefficient $\beta_2$) is not statistically significant for the parameters derived from the Normal density fit. 
However, all the other parameters in the Normal and SN cases are statistically significant for modelling stellar activity. 
By analyzing the values of the coefficient of determination, $R^2$, we see that the model for SN mean RV is able to capture the highest percentage of variability in the estimated RV. 
This is not a surprising result since the three different RV estimates have the same RV residual rms after correction for activity, but before correction, SN mean RV had the largest RV rms (see Fig~\ref{fig:alphacent:correctionRV}).

{  When looking at the results discussed in this section, it is likely that the activity signal of Alpha Centauri B is due to faculae. 
Like observed for the simulated facula in Sec.~\ref{sec:soap.faculae}, the amplitude of the activity signal is slightly stronger for SN mean RV
than for N mean RV or SN median RV; the amplitude of the two latest being comparable. In addition, when applying the proposed correction for
activity in the case of the Alpha Centauri B data, the interaction term is significant, which was only the case for the simulated facula in 
Sec.~\ref{sec:soap.faculae}. Those are arguments strengthen the findings of \citet{Dumusque-2014c} that also found evidence for faculae
dominating the RV stellar signal of Alpha Centauri B.}
%Another feature that suggests the presence of faculae is the strong positive measured correlation between $\gamma$ and SN mean RV and  between SN FWHM and SN mean RV, as displayed in Fig.~\ref{fig:alphacent:corrPlot}; 
%as we saw in Sec.~\ref{sec:soap.spot} a spot induces a negative correlation between $\gamma$ and SN mean RV and weak correlations are measured between SN FWHM and SN mean RV.
%When using SN mean RV, it is possible to observe more variations than the ones measured by the Normal fitting. This happens because the mean of the SN is more sensitive to stellar activity. In fact, because the SN includes an asymmetry parameter, SN mean RV gets more shifted in the direction of the asymmetry induced by stellar activity. On the other hand, when using SN median RV, smaller variations in RV are caused by changes in the asymmetry of the CCF, because this second location parameter is a more robust indicator than the mean. The bottom plot of Fig.~\ref{fig:alphacent:diff:RV} captures this aspect. Both indicators can be used to capture and summarise the different information available in the CCF, as will be shown in the remainder of this work.
%%

\begin{table}
\begin{center}
\caption{P-values for the different coefficients used in Eq.~\eqref{eq:RV:correction} for the correction from stellar activity in Alpha Centauri B data. All the variables corresponding to the Normal or SN parameters are statistically significant, except for the asymmetry of the CCF when using the BIS SPAN with N mean RV.
The estimated $R^2$ show that the proposed linear combination explains the most variability in RVs due to the stellar activity when the RVs are estimated with the SN mean RV.}
\label{table:alphacent.test}
\begin{tabular}{|c|c|c|c|}
\hline
Parameter          & N mean RV         &   SN mean RV &   SN median RV \\
\hline
$\beta_{0}$            &    $0.49$    & $0.90 $  & $0.027$ \\
\hline
$\beta_{1}$            &    $2.22e-16$    & $2.22e-16 $  & $2.22e-16$ \\
\hline
$\beta_{2}$            &     $0.33$   & $2.22e-16 $ & $1.23e-11$\\
\hline
$\beta_{3}$            &     $ 2.22e-16$   &  $2.22e-16 $  & $ 2.22e-16$\\
\hline
$\beta_{4}$            &     $2.22e-16$   &  $2.22e-16 $ & $ 2.22e-16 $\\
\hline
$R^{2}$      &     $0.57$    &  $0.78$ & $0.66$  \\
\hline
\end{tabular}
\end{center}
\end{table}

%Both the proposed indicators coming from the SN density have advantages and limits: SN mean RV better catches changes in the asymmetry of the CCF but the resulting set of RVs ends up being contaminated by those spurious shifts caused by stellar activity that have been shortly presented in Sec. \ref{intro}. When using SN median RV, the final set of RVs is less affected by those spurious shifts caused by stellar activity, but at the same time this indicator is not able to catch as well as SN mean RV changes in the shape and in the width of the CCF. Once corrected from stellar activity using Eq.~\eqref{eq:RV:correction}, the results are comparable. Anyway, both SN mean RV and SN median RV are useful to catch different aspects of the CCF and our suggestion is to use SN mean RV when interested in retrieving information about changes in the shape and/or the width of the CCF. In order to provide a set of RVs containing the smallest amount of spurious contamination imputable to stellar activity (i.e. before to run Eq.~\eqref{eq:RV:correction}), our suggestion is to use instead SN median RV. 

%-----------------------------------------------------------------------------------------------------------------------------------------------
\subsection{Comparison for HD192310, HD10700, HD215152 and Corot-7 of the different CCF parameters derived with the Normal and the Skew Normal} \label{sec:real_data_other_stars}

In the previous section we evaluated for Alpha Centauri B the improvement obtained by the SN parameters compared to the Normal parameters and the BIS SPAN. We carryout similar analyses for four other main sequence stars: HD192310 \citep[K2V,][]{Pepe-2011}, HD10700 \citep[G8V,][]{Feng:2017ac}, HD215152 \citep[K3V,][]{Delisle:2018aa} and finally Corot-7 \citep[K0V,][]{Haywood-2014}. The same correlation and residual plots displayed in the previous section for Alpha Centauri B can be found for those new four stars in Appendix~\ref{appendix}.

The correlations between the parameters of these additional stars are similar to those obtained for Alpha Centauri B. The correlation between $\gamma$ and SN mean RV is the strongest among all the asymmetry-RV correlations. Between the width parameters and the estimated RV, the strongest correlation often is between SN FWHM and SN mean RV. 
However, there is one exception in the case of HD10700 where the Pearson correlation coefficient between FWHM and N mean RV is equal to $R=0.53$, while it is $R=0.42$ between SN FWHM and SN mean RV, and $R=0.5$ between SN FWHM and SN median RV. {  Like in the case of Alpha Centauri B, positive correlations are always observed between the asymmetry and barycentre of the CCF. This cannot be explained by SOAP 2.0 and to our knowledge by other stellar activity simulators, showing the limit of these tools.}

Except for the special case discussed above for HD10700, the analyses of those four stars, in addition to the analyses on Alpha Centauri B, show that the parameters derived when using a SN density are generally more sensitive to activity.  
Therefore using the SN parameters, and in particular estimating RV using SN mean RV, can result in better detection of stellar activity than the Normal parameters. 
More specifically, this is the case for the evaluation of the asymmetry-RV correlations for Alpha Centauri B, HD10700, HD215152, HD192310 and Corot-7, and the width-RV correlation for Alpha Centauri B, HD215152, HD192310 and Corot-7 (see Appendix \ref{appendix}).

When correcting for stellar activity for  Alpha Centauri B, although the uncorrected RV rms was larger for SN mean RV (compared to the RVs obtained using N mean RV), once corrected for activity using the new model proposed in Sec.~\ref{sec:31}, both RVs estimates had similar residuals. 
For HD10700, HD215152, and HD192310, the proposed and usual models were giving similar RV residual rms.
%%%We also observe slight differences in the rms between the new proposed correction and the usual correction, which uses only a linear combination of the width and the asymmetry of the CCF.  
However, for Corot-7, the new correction is able to provide RV residual rms 23\,\cms\,smaller than the one obtained with the usual correction.

%-----------------------------------------------------------------------------------------------------------------------------------------------
\subsection{Detection limits when using the estimated RVs from the Normal or the Skew Normal models} \label{sec:detect_limits}

In the previous section, we saw that the estimated RV resulted in different amplitudes when considering a SN or a Normal density, especially when using SN mean RV. 
However, once corrected for stellar activity using the linear combination presented in Eq.~\eqref{eq:RV:correction}, as shown in the bottom plots of Fig~\ref{fig:alphacent:correctionRV}, the rms of the residuals are essentially the same for all three approaches.
In this section, we investigate the ability of the three different RV estimators (N mean RV, SN mean RV, and SN median RV) to detect planetary signals among stellar activity, and also compare them when using the usual stellar activity correction with the proposed stellar activity model of Eq.~\eqref{eq:RV:correction}.
To carryout this test, the minimum detected amplitude of an injected planetary signal is estimated at different orbital periods when considering data affected by stellar activity.

In order to obtain CCFs affected by realistic stellar activity signals, the CCFs from Alpha Centauri B used previously were considered. 
To simulate a planetary signal, the CCFs were blue- or red-shifted with the desired amplitude, period, and phase.
Several RV data sets with the same stellar signal, but different planetary signals were generated using parameters corresponding to the following grid:
\begin{itemize}
\item period of 3, 5, 7, 9, 11, 15, 20, 25, and 30 days,
\item amplitude from 0.5 to 3 \ms\, by steps of 0.05\,\ms,
\item 10 different phases, evenly sampled between 0 and 2$\pi$.
\end{itemize}

For each of the 4590 simulations we computed the three estimates of RV, namely N mean RV, SN mean RV and SN median RV. 
On each set of RV estimates, we performed an analysis similar to Sec.~\ref{sec:soap.spot.planet}, i.e. fitting the activity signal using Eq.~\eqref{eq:RV:correction}
or the usual correction along with a circular planetary signal (see Eq.~\eqref{eq:RV:correction.overall}). Because of the non-linearity of the model that includes a 
planet, a non-linear least squares algorithm was used for the fit \citep[][]{levenberg1944method,marquardt1963algorithm,teunissen1990nonlinear}. Such a model requires
initial conditions close to the real solution, otherwise the algorithm can converge to a local minimum. Because our goal here is to compare the planetary detection limits
using the three different RV estimates and the two different activity models proposed, and not to discuss what is the best method to explore the parameter space, we 
initialised the minimisation algorithm to the real period of the planetary signal injected to avoid getting stuck in a local minimum. We also selected as initial amplitude the peak-to-peak amplitude of the estimated RVs.
The argument of periapsis $t_0$ was initialised to the time when the RV was crossing 0 since we use a sinusoidal function to fit the planetary signal (see Eq.~\eqref{eq:RV:correction.planet}).

\begin{figure}[!h]
\begin{center}
\includegraphics[height = 2.6in]{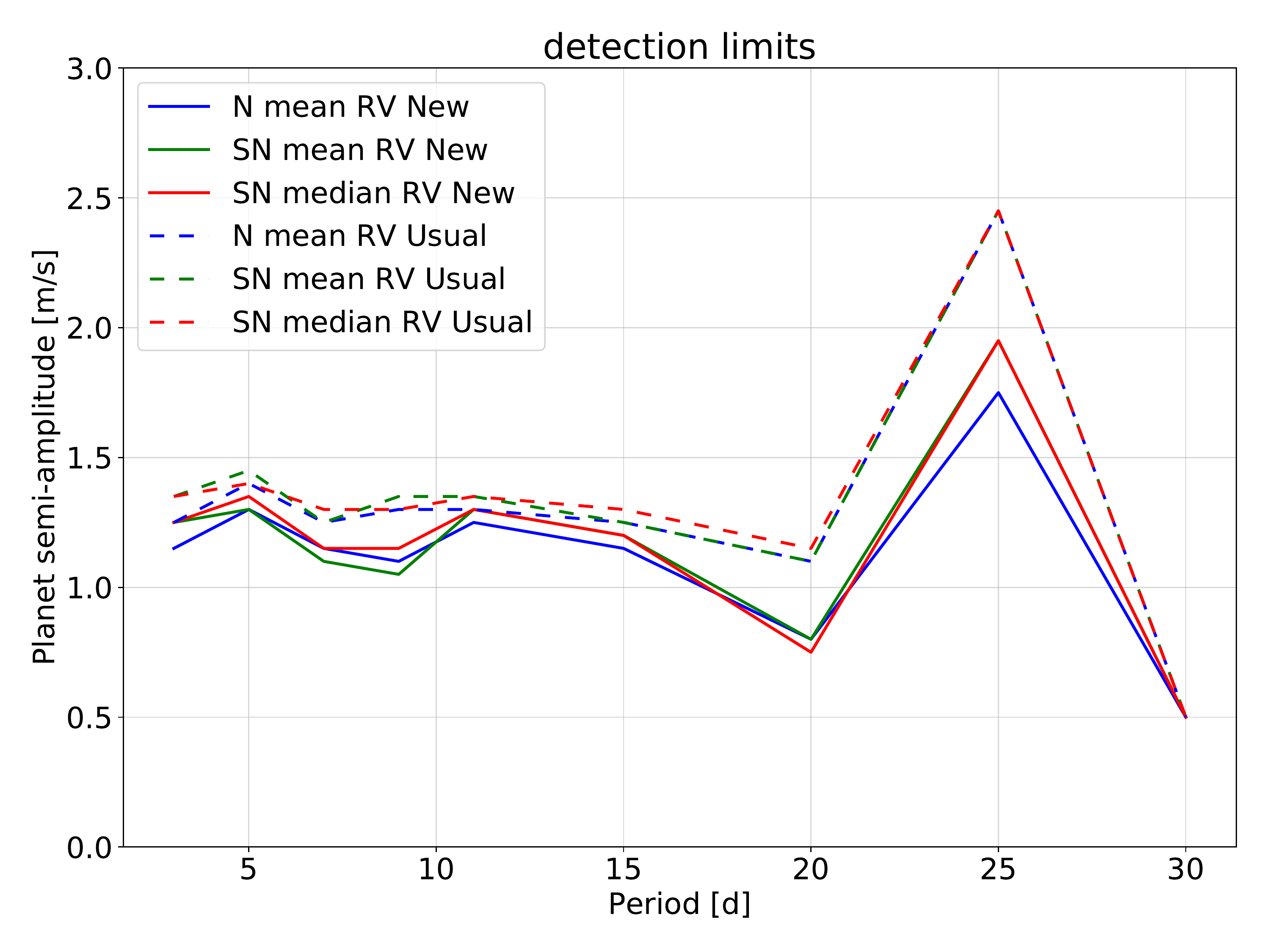} 
   \caption{Detection limits of planetary signals once the stellar activity signal is removed from the raw RVs using the model proposed in Eq.~\eqref{eq:RV:correction} (solid lines) or the usual correction based on $RV_{\text{activity}}=\beta_0+\beta_1 \gamma + \beta_2 \text{SN FWHM}$ for the SN fit and on $RV_{\text{activity}}=\beta_0+\beta_1 \text{BIS SPAN} + \beta_2 \text{FWHM}$ for the normal fit (dashed lines). The correction for stellar activity based on Eq.~\eqref{eq:RV:correction} improves on average the detection limit by 12\% and the different RV estimators have similar detection limits.}
   \label{fig:detection_limits}
\end{center}
\end{figure}

Once the parameters involved in Eq.~\eqref{eq:RV:correction.overall} were estimated,  signals in the residuals, defined as $RV - RV_{\text{activity}}$, were analyzed using a Generalized Lomb-Scargle periodogram \citep[][]{Lomb-1976a, Scargle-1982, Zechmeister-2009}. 
If a signal with a P-value\footnote{The P-values were estimated using a bootstrap procedure.} smaller than 1\% had a period compatible with the injected planetary period within an error budget of 20\%, the signal was considered significant and the corresponding planet considered detected.  
For each period considered, we searched for the minimum amplitude at which at least 80\% of the planets with different phases were detected.
This minimum amplitude detected as a function of period is shown in Fig.~\ref{fig:detection_limits} for the three different RV estimates (N mean RV, SN mean RV, and SN median RV) when using the new stellar activity correction proposed in this paper (see Eq.~\eqref{eq:RV:correction}), and when using the usual activity correction.
We can see that the new correction for stellar activity based on Eq.~\eqref{eq:RV:correction} improves the detection limit of the exoplanet by 12\% on average compared to the usual approach, and the three estimators of RV give similar detection limits.
These results therefore suggest that any of the RV estimators can be used when searching for a planetary signal in RV data contaminated by stellar activity, and using our new model to account for stellar activity allows to detect planetary signals with a slightly smaller amplitude that the usual correction that uses only a linear correlation with the FWHM and BIS SPAN.

%Although the detection limits remain relatively constant for the periods  $\leq$20 days, there is a notable increase in the detection limit at 25 days. 
%This is probably due to interaction between the planetary signal and the stellar activity, however 
%
%the fact that the simulated planets at 25 days are close to the first harmonic or to the rotational period of the star \citep[36.7 days,][]{Dumusque-2012}.
%However this reasoning 
%This is likely caused by the fact that the periods of the simulated planets at 25 days are close to the first harmonic or to the rotational period of the star \citep[36.7 days,][]{Dumusque-2012} and therefore close to the semi-periodicity of the stellar activity signal. \jessi{wouldn't a period of 30 be considered closer to the rotation period and so should also be higher based on this reasoning?}

%-----------------------------------------------------------------------------------------------------------------------------------------------
\section{Estimation of standard errors for the CCF parameters} \label{sec:5}

In this section, we investigate how the photon noise influences the CCF parameters derived either by a Normal density or a SN density fit. 
Because a CCF is obtained from a cross-correlation, each point of a CCF is correlated with the other points. Therefore, we cannot simply vary each point in the CCF by their respective error bars and then recalculate the best SN or Normal density fit to see how the CCF noise influences the estimation of the parameters of interest (i.e., N mean RV, SN mean RV, SN median RV, FWHM, SN FWHM, BIS SPAN and $\gamma$). 
Instead, we go to the individual spectrum where each individual points can be considered independent from the others.
The standard error on each point of a spectrum is given by the photon noise, which follows a Poisson distribution and is therefore estimated by taking the square-root of the measured flux.

The following method was carried out in order to estimate the error bars on the different parameters derived from the CCF. 
We first modify the values of all the points in the spectrum given their respective error bars. 
To do so random Gaussian noise with standard deviation the square-root of the flux was added across each spectrum. 
The CCF was calculated using this spectrum according to the method presented in \citet{Pepe-2002a}, then fit by either a Normal or SN density with the parameters recorded. 
This process was repeated a hundred times in order to obtain a distribution for each CCF parameter, and the standard deviations of the resulting distributions provide estimates of the standard errors for the CCF parameters.

\begin{figure*}[htbp]
\begin{center}
\includegraphics[height = 6in]{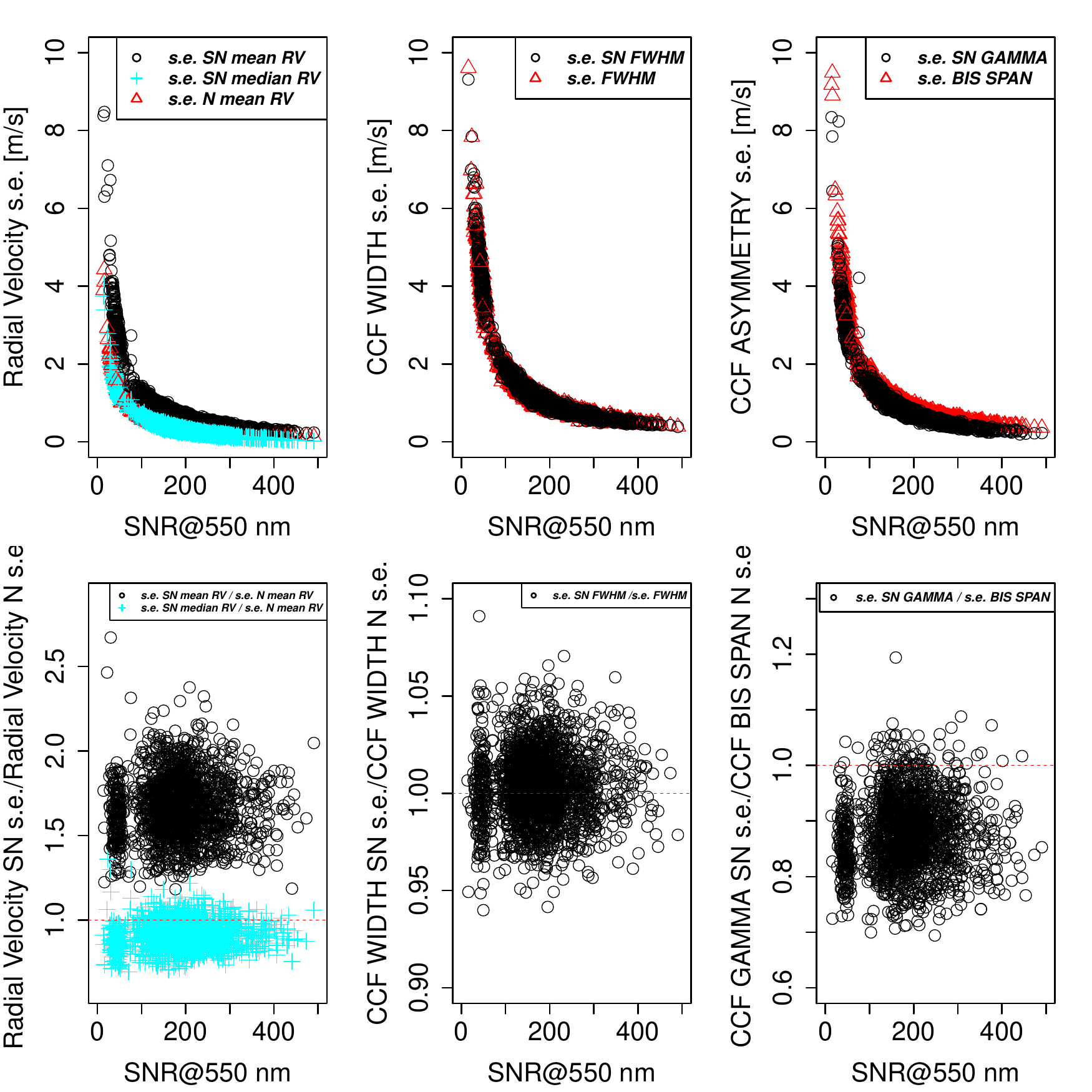} 
   \caption{Results of the bootstrap analyses on the stars HD215152, HD192310 and Corot-7. (top) Comparison between the standard errors from the bootstrap analysis of the estimated RVs, FWHM, and asymmetry parameters using the SN fit and the common strategy (Normal fit and BIS SPAN). (bottom) Ratio between the standard errors estimated on the parameters derived from the common strategy and the corresponding standard errors estimated on the parameters derived from the SN fit. When using SN mean RV (black circles), the standard errors are in average $60\%$ larger than the standard errors of N mean RV (red triangles). However, the standard errors for SN median RV (cyan pluses) are on average $10\%$ smaller than the standard errors coming from the N mean RV. 
   The use of the asymmetry SN parameter $\gamma$ leads to standard errors in average $15\%$ smaller than the standard errors related to the BIS SPAN. Note that for asymmetry, the error in BIS SPAN is in \ms. To be able to compare the errors in $\gamma$ and BIS SPAN, we multiplied the error in $\gamma$ by the slope of the correlation between $\gamma$ and BIS SPAN.}
   \label{fig:se}
\end{center}
\end{figure*}

The standard errors were computed for each CCF parameter for the HARPS measurements of HD215152, HD192310 and Corot-7. 
These three stars include measurements that cover the range of S/N measured at 550 nm (S/N550) from 10 to 500, which represent the very low S/N limit and the saturation limit of the HARPS detector, respectively. 
HD10700 and Alpha Centauri B were not included because they have a large number of measurements, which would require a substantial computational effort.
The variation of the noise for each CCF parameter as a function of S/N550 is displayed in Fig.~\ref{fig:se}.
The top row shows the standard errors of the three different estimated RVs, the width, and the asymmetry estimates. 
Note that because BIS SPAN and $\gamma$ do not have the same units, the estimated slopes of the correlation between those two parameters to transform $\gamma$ in \ms were used (see Fig.~\ref{fig:alphacent:corr.gamma} and Table~\ref{table:summaryStars} for the value of the slope for each star). 
The bottom row shows the ratio between the standard errors measured when using the SN parameters and the Normal parameters. Values smaller (larger) than one will imply that standard errors from the SN parameters are more (less) precise than the Normal parameters.

{  The standard errors for the different RV estimates all appear to follow a similar exponential decay as a function of S/N, even though the measurements are from three different stars. This suggests that, for the considered stars, the precision in RV is mainly driven by the S/N of the analyzed spectra. As shown in \citet{Bouchy:2005aa}, the RV precision is proportional to the S/N, the FWHM of the CCF and its contrast. In our case all three studied stars are main sequence K-dwarfs, which imply that their CCF FWHM and contrast are similar and explains why the RV precision is driven by the S/N only.}

When comparing the three different estimates for the RV, SN mean RV has standard errors that are 60\% larger than N mean RV. However, SN median RV gives errors 10\% more precise than N mean RV. 
The parameters describing the width of the CCF, FWHM and SN FWHM, have comparable standard errors. 
Finally, for the asymmetry parameters, $\gamma$ has standard errors that are 15\% more precise than BIS SPAN. 
In conclusion, when fitting a SN density to the CCF and when using SN median RV as the RV estimate, we are able to improve the precision on the estimated RV by 10\%. Using the SN density, we are also able to improve by 15\% the precision on the estimated asymmetry parameter of the CCF.  However, SN mean RV should not be use to derive precise RV estimates
%%, except perhaps in specific conditions described below, 
as the precision on this parameter is 60\% worse than the precision on the RVs derived from N mean RV.

\section{Discussion} \label{sec:discu}

When fitting a SN density shape to the CCF, parameters used to estimate the RV, defined as the CCF barycenter, the amplitude, sometimes called the CCF contrast, the width and the asymmetry of the CCF can all be estimated in a single model framework.
For the estimation of the RV, we investigated the use of the mean and the median of the SN density. 
The width is derived using the variance of the SN density (SN FWHM$=2\sqrt{2\text{ln}(2)\sigma^2}$) and the asymmetry by using $\gamma$, skewness parameter of the SN density.

To evaluate the performance of the proposed SN framework, tests on both simulated and real data were carried out and compared to the commonly employed approach of fitting a Normal density shape to the CCF to get access to the RV and the FWHM, and then separately deriving the BIS SPAN to estimate the CCF asymmetry.
The simulated CCFs were generated using the SOAP 2.0 code, which can simulate activity signals induced by a spot or a facula on a solar-like star. {  To simulate realistic data, we considered a S/N of 100, which is typical for high-precision RV observations.}

The results of the simulation study suggest that {  at least one of the parameters} derived from the SN density fit is equally or more sensitive to activity than the parameters obtained by the usual Normal method, making this or these parameters more useful indicators of activity. {  Sensitivity was measured by the strength of the correlation between the different SN or Normal parameters. In the case of a spot, the strongest correlations are found between $\gamma$ and SN median RV and between BIS SPAN and N mean RV, therefore making the SN parameters equally sensitive to activity. For the facula case, the strongest correlation is between SN FWHM and SN mean RV with a correlation coefficient of R=0.92. The correlation between the parameters derived from the Normal fit are much weaker, with correlation coefficients  between BIS SPAN and N mean RV of R=-0.61 and between FWHM and N mean RV of R=0.47.}
The SN parameters continued to have stronger correlations than the Normal ones in the setting where a planetary signal was added the SOAP 2.0 with a single spot.

{  Looking at real data, we arrive to a similar conclusion that the SN parameters are more sensitive to activity. While real data confirms that the correlation between CCF width and barycentre is always stronger for the SN parameters, they also show that it is the case for the correlations between the CCF asymmetry and barycentre. In this later case, all stars studied in this paper show a positive correlation, which cannot be explained by SOAP 2.0, as the spot and facula simulations only show negative correlations between the CCF asymmetry and barycentre. This discrepancy between simulations and real data could be} due to the fact that SOAP 2.0 uses the spectrum of a spot as input to model the activity induced by a facula.
Because the temperature between a spot and a facula is significantly different, their spectra should be different. 
Additionally, there are expected to be multiple active regions on a star at different locations in longitude and latitude, while the SOAP 2.0 data used in the simulation study included only a single active region on the equator {  in order to isolate the effects of those active regions}. 
%\xavier{The fact that we also find significant differences between the simulated and real data for the correlations between BIS SPAN and N mean RV and between FWHM and N mean RV shows that the SOAP 2.0 simulation studied here are too simple to explain the complexity of stellar activity in real RV observations.}
%\jessi{What are these differences for Normal case?  The specific difference for the SN is the strength of the correlations with the SN mean RV compared to the SN median RV.  When I look at the SOAP vs. real data for the Normal, I do not notice anything specifically different other than different patterns, but that is not surprising.}
%\umberto{I suggest to comment only about the differences for the SN case, since for the Normal I do not see differences between the analyses with SOAP2.0 and real stars.}

In the real real cases, the parameters derived from the SN are always more sensitive to activity than the parameters derived from the Normal.  
{  There is only one exception in the case of the width-barycentre correlation for HD10700, however, the difference in correlation between the SN and Normal parameters is rather small, from R=0.42 to 0.53, respectively.}
Also, the correlation between the asymmetry and SN mean RV is consistently stronger than the parametrization of the CCF presented in \citet{Boisse-2011} and \citet{Figueira-2013}. 
Because the apparent RV signal induced by activity  
results in a stronger correlation with the SN parameters
than between the apparent RV signal and the FWHM of the CCF or its BIS SPAN, this suggests the SN model of the CCF can lead to a better understanding of the spurious variations in RV caused by stellar activity.

Considering the different RV estimates of the real data, the amplitude of stellar activity tends to be largest for SN mean RV, followed by SN median RV and N mean RV, which behave similarly. 
This implies that the mean of the SN density appears to be more sensitive to variation in the CCF shape than the median of the SN or the mean of the Normal.  
%This is not surprising as the SN density is able to model the asymmetry of the CCF, what the Normal density does not. \jessi{The previous sentence does not follow from the points made in the paragraph.}

Having an estimator of RV that is more sensitive to stellar activity, such as SN mean RV, can also help to better probe stellar rotational periods or to better understand the covariance of stellar signals when fitting a Gaussian Process to the RVs \citep[e.g.][]{Faria-2016a, Haywood-2014}. 
We saw that the SN mean RV estimator is 60\% noisier than the N mean RV estimator. 
%\xavier{This is of course a problem for photon-noise limited observations.} 
However, when looking at bright
stars like $\alpha$\,Centauri\,B or HD10700, increasing the photon noise by 60\% will not have a significant impact
on the RV precision as the instrumental noise will dominate the data. 
Therefore, for bright targets, stellar activity 
can be better characterized by using SN mean RV as this RV estimate is more sensitive to it.
%
%This is not necessarily a negative aspect. 
%For example, in the case where photon-noise is not dominant, stellar activity can be better characterized if its effect can be amplified by measuring the SN mean RV, even if the uncertainties related to this parameter are increased by 60\%.
%\jessi{I'm not sure I agree with the previous statements - more uncertainty means you would be less certain if the RV estimate is due to actual stellar activity or just photon noise.  It is also stated throughout that precise estimates are needed (e.g. in the conclusion)}
%\umberto{Xavier, can you please develop a little bit this statement? If I understood correctly, since the correlation between the width and the asymmetry of the CCF with SN mean RV is almost always the strongest, if the star is particularly active we are willing to ``sacrifice'' the precision of the RV if we can end up with a clear indication of strong activity by the star. Is it correct?}

We also propose a new model to correct the estimated RV data for stellar activity signals. 
Generally, when fitting for planetary signals, it is common to use a model composed of one or several Keplerian signals to account for the planets, in addition to a linear combination of the FWHM and BIS SPAN to account for stellar activity signals. 
The proposed model adds a term to the linear model to account for the amplitude of the CCF and an interaction term between the estimated asymmetry and the width parameters. 
Using the simulated data from SOAP 2.0, this new model reduces the effect of the stellar activity signal by factors of about 2 and 3.5 over the usual model, respectively, for the facula and the spot.

Even if the different RV estimators {  derived by the Normal or SN fit} result in different amplitudes, once the proposed correction for stellar activity is applied, the residuals of the model have similar rms.
When comparing the activity correction proposed in this paper with the usual correction that only uses a linear combination of the CCF asymmetry and width, for the simulations based on the presence of a facula or a spot the new proposed correction almost entirely explains the spurious variations in RV. However, when moving to real data, there is just a slight improvement by using the proposed correction function for stellar activity. 
Additional analysis can be performed for new datasets to see if {  certain components of the model proposed here are not relevant and, therefore, could be removed.}

A test was carried out to see if some RV estimates were better at finding planets in RV data affected by observed stellar signals. The new correction based on Eq.~\eqref{eq:RV:correction} proposed in this paper to mitigate the effect from stellar activity slightly improves the detection limit with respect to the usual one based on $RV_{\text{activity}}=\beta_0+\beta_1 \gamma + \beta_2 \text{SN FWHM}$ for the SN fit and on $RV_{\text{activity}}=\beta_0+\beta_1 \text{BIS SPAN} + \beta_2 \text{FWHM}$ for the Normal fit.
Concerning the definition of the RV using the SN or the Normal fit, all three of the different RV estimators give similar detection limits. Therefore it seems that any of the RV estimators can be used to search for planetary signals.

Finally, we investigated the precision of each of the SN and Normal parameters including the BIS SPAN. It turns out that SN mean RV should not be used to get precise RVs as the standard errors on this parameter is 60\% greater than for N mean RV. 
However, SN median RV is 10\% more precise than N mean RV. 
Regarding the asymmetry estimates, we observe that $\gamma$ has a precision 15\% better than the BIS SPAN.

%Finally, we also encourage the use of bootstrapping to estimate more realistic errors on the different parameters of the Normal or SN fitted to the CCF, mainly in the low S/N regime where a gain of 50\% can be reached. This takes significantly more time, but note that 100 bootstrapped dataset are enough to get a good estimation of errors.

%%%%%%%%%%%%%%%%%%%%%%%%%%%%%%%%%%%%%%%%%%%%%%%%%
\section{Conclusion} \label{sec:conclu}

When searching for low-mass exoplanets using the RV technique, it is necessary to retrieve precise estimates of the RV and also to account for variations induced by stellar activity in order to avoid false detections.  
Stellar activity such as spots and faculae can lead to shape variations in the spectra features, which then results in shape variations of the CCF.
The correlations between the width or asymmetry of the CCF and the estimated RV are commonly used as a way to detect if the RVs are affected by stellar activity signals.   
Because the presence of real planets would result in only a shift in the CCF (not a change of its shape), strong correlations between the shape features of the CCF and the estimated RVs suggest that stellar activity may be present.

In this paper, a new approach for quantifying shape changes in the CCF is proposed using the SN density, which can be used to estimate with a single fit the RV, the width and the skewness of the CCF. 
This new method is compared to a commonly used method based on a Normal density fit to the CCF.  The mean of the Normal density is used as the estimated RV and the FWHM estimates the width of the CCF.  Because the Normal density does not have any skewness, another method is necessary to estimate the asymmetry of the CCF, such as the often employed BIS SPAN.
In addition, the proposed SN approach is compared to other parameterizations of the CCF asymmetry that have been shown to be sensitive to activity signals \citep[][]{Boisse-2011,Figueira-2013}.

In the different tests carried out for this work, the SN parameters performed at least as well as, and most of the times better than the parameters from the Normal approach and the BIS SPAN.
The SN parameters $\gamma$, SN FWHM, and SN mean RV consistently had stronger correlation than those between any of the parameters derived by the Normal and the BIS SPAN, or the different asymmetry parametrizations presented in \citet{Boisse-2011} and \citet{Figueira-2013}. 
This suggests the SN parameters may be better at probing stellar activity signals than the other methods. 
In addition, the uncertainties measured on SN median RV and $\gamma$ are, respectively, 10\% and 15\% smaller than the corresponding uncertainties on N mean RV and BIS SPAN, though SN mean RV had uncertainties 60\% greater than N mean RV.

Because of the advantages of using the proposed SN approach over the commonly employed approach based on the Normal density fit to the CCF and the BIS SPAN or the asymmetry parameters described in \citet{Boisse-2011} and \citet{Figueira-2013}, a SN density model for the CCF may be more useful for detecting stellar activity than the previously proposed parametrizations.
Correlations between $\gamma$ and SN mean RV, and between the width and SN mean RV can be used to probe stellar activity signals in RV data, and SN median RV can be used to estimate RV.
We also proposed a new model to correct the estimated RV data for stellar activity signals, by using the amplitude of the CCF and an interaction term between the estimated asymmetry and the width parameters. Using simulated data from SOAP 2.0, { {this new proposed correction reduces the effect of the stellar activity signal by an additional 14.5 -15 \% and 5.3-5.8 \%}} over the usual model, respectively, for facula and spot. When applying this model on real data, we observe that planetary detection limits are improved by a non-negligible 12\%.

\section{Acknowledgements}

The authors thank Yale's Center for Research Computing for their help and resources with some of the computational aspects of this work.
XD is grateful to The Branco Weiss Fellowship--Society in Science for its financial support.
JCK was partially supported by the National Science Foundation under Grant AST 1616086 and by the National Aeronautics and Space Administration under grant 80NSSC18K0443.
US was partially supported by Fondazione CARIPARO and thanks the IT-University of Helsinki for the computational resources provided to execute part of the analyses of the present work.
The authors are grateful to all technical and scientific collaborators of the HARPS Consortium, ESO Headquarters and ESO La Silla who have contributed with their extraordinary passion and valuable work to the success of the HARPS project.

%-----------------------------------------------------------------------------------------------------------------------------------------------
%\section{Appendix}
\appendix
\section{Appendix} \label{appendix}

In this Appendix, a similar analysis as the one presented in Sec.~\ref{sec:4} is discussed for four main-sequence stars: HD192310 \citep[K2V,][]{Pepe-2011}, HD10700 \citep[G8V,][]{Feng:2017ac}, HD215152 \citep[K3V,][]{Delisle:2018aa} and finally Corot-7 \citep[K0V,][]{Haywood-2014}. The latest HARPS data for these stars can be found on the ESO archive.

Table~\ref{table:summaryStars} summarizes the results obtained by the SN and Normal density models of the CCF. 
These results are consistent with those from the analysis of Alpha Centauri B. 
The correlation between $\gamma$ and SN mean RV is stronger than the correlation between the BIS SPAN and N mean RV or between the asymmetry parameters derived in \citet{Boisse-2009} and \citet{Figueira-2013} and N mean RV for all the considered stars. 
The correlation between SN FWHM and SN mean RV is stronger than the correlation between FWHM and N mean RV for three of the four stars.  
Also for all these stars, the originally estimated RVs were corrected from spurious variations caused by stellar activity using Eq. \ref{eq:RV:correction} and Fig.~\ref{fig:HD192310:correctionRV}, \ref{fig:HD10700:correctionRV}, \ref{fig:HD215152:correctionRV}, and \ref{fig:Corot-7:correctionRV}, show the corrected RVs. 
Once corrected from stellar activity, the Normal and SN residuals are comparable for the stars $\text{HD}192310$, $\text{HD}10700$ and $\text{HD}215152$.
However, the rms of the residuals for Corot-7 are 0.23\,\ms\,lower for the SN model than the Normal model.  %\xavier{$\text{HD}215152$ and Corot-7 have in average a lower S/N at 550 nm than HD10700 and HD192310, 273, 207, 141 and 44, respectively.}
The average S/N at 550 nm for the stars \text{HD}10700, \text{HD}192310, $\text{HD}215152$ and Corot-7 are, respectively, 273, 207, 141, and 44.
Corot-7 has therefore on average a much lower S/N at 550 nm than the others stars, which could be a potential explanation for this small improvement. Additional tests should be performed to confirm this statement.

\small{
\begin{table*}[!t]
\caption{Notable correlations between the asymmetry or the FWHM parameters and the estimated RVs for four stars: $\text{HD}192310$,  $\text{HD}10700$, $\text{HD}215152$ and $\text{Corot }7$. The complete results of the analyses of the correlations for the four stars are presented in Fig. \ref{fig:Gliese785:corrPlot}, \ref{fig:Tau:corrPlot}, \ref{fig:HD215152:corrPlot}, and \ref{fig:Corot7:corrPlot}.}
\label{table:summaryStars}
\begin{center}
\scalebox{0.9}{
\begin{tabular}{ccccccccc}
\textbf{Star}          &\textbf{ \# CCFs}  &   \textbf{$\text{R}(\text{SN }\gamma, \text{Bis-Span})$} & \textbf{$\text{slope}(\text{SN }\gamma, \text{Bis-Span})$} &   \textbf{$\text{R}(\text{SN }\gamma, \text{SN mean RV})$}\\
\hline
\hline
 $\text{HD}192310  $          &    $1577$    & $0.888$ & $0.000786$ & $0.669 (0.64; 0.695)$\\
\hline
 $\text{HD}10700 $            &    $7928$    & $0.78$ & $0.000604$ & $0.322 (0.302; 0.342)$\\
\hline
 $\text{HD}215152 $          &     $273$   &  $ 0.763$ & $0.000794$ & $0.571 (0.485; 0.646)$ \\
\hline
 $\text{Corot }7$     &     $173$    &  $0.814$  & $0.000607$ & $0.561 (0.450; 0.656)$  \\
\hline
\rule{0pt}{2ex}    & & & &\\
\textbf{Star}          &\textbf{$\text{R}(\text{Bis-Span}, \text{N mean RV})$} & \textbf{$\text{R}(\text{FIG BiGaussian}, \text{N mean RV})$} & \textbf{$\text{R}(\text{SN FWHM}, \text{SN mean RV})$}  & \textbf{$\text{R}(\text{FWHM}, \text{N mean RV})$} \\
\hline
\hline
 $\text{HD}192310  $& $0.329 (0.285; 0.373)$  & $-0.333 (-0.376; -0.289)$ & $0.666 (0.637; 0.692)$ & $0.476 (0.4367; 0.514)$\\
\hline
 $\text{HD}10700 $ & $-0.073 (-0.095; -0.0051)$ & $0.127 (0.105; 0.148)$ & $0.421 (0.403; 0.439)$ & $0.529 (0.513; 0.545)$ \\
\hline
 $\text{HD}215152 $& $-0.067 (-0.184; 0.052)$  & $0.269 (0.155; 0.376)$ & $0.210 (0.094; 0.321)$ & $-0.138 (-0.253; -0.020)$ \\
\hline
 $\text{Corot }7$ & $0.092 (-0.058; 0.238)$ & $-0.335 (-0.228; -0.082)$ & $-0.709 (0.626;0.776)$ & $0.595 (0.489; 0.683)$ \\
\hline
\rule{0pt}{2ex}    & & & &\\
\end{tabular}}
\end{center}
\end{table*}
}

%HD192310
\begin{figure*}[htbp]
\begin{center}
\includegraphics[height = 6in]{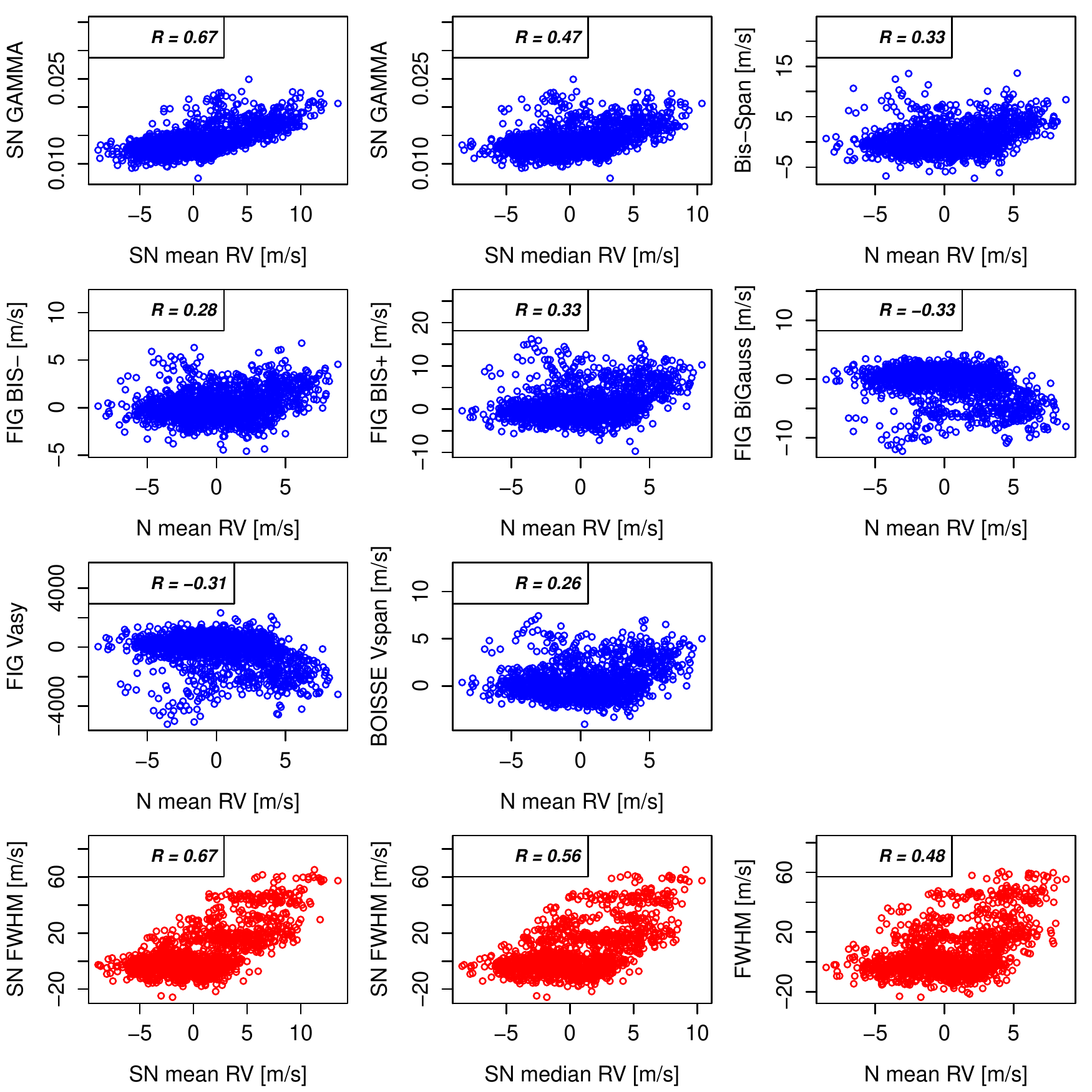} 
   \caption{(top three rows) Correlations between the asymmetry parameters and their corresponding RVs for $\text{HD}192310$.  
   (bottom row) Correlations between the FWHM and the estimated RVs. 
   The correlations are consistently stronger when using parameters derived from the SN than the Normal. The estimated $R$ are all statistically significant.}   
   \label{fig:Gliese785:corrPlot}
\end{center}
\end{figure*}

\begin{figure*} 
\begin{center}
\includegraphics[height = 6in]{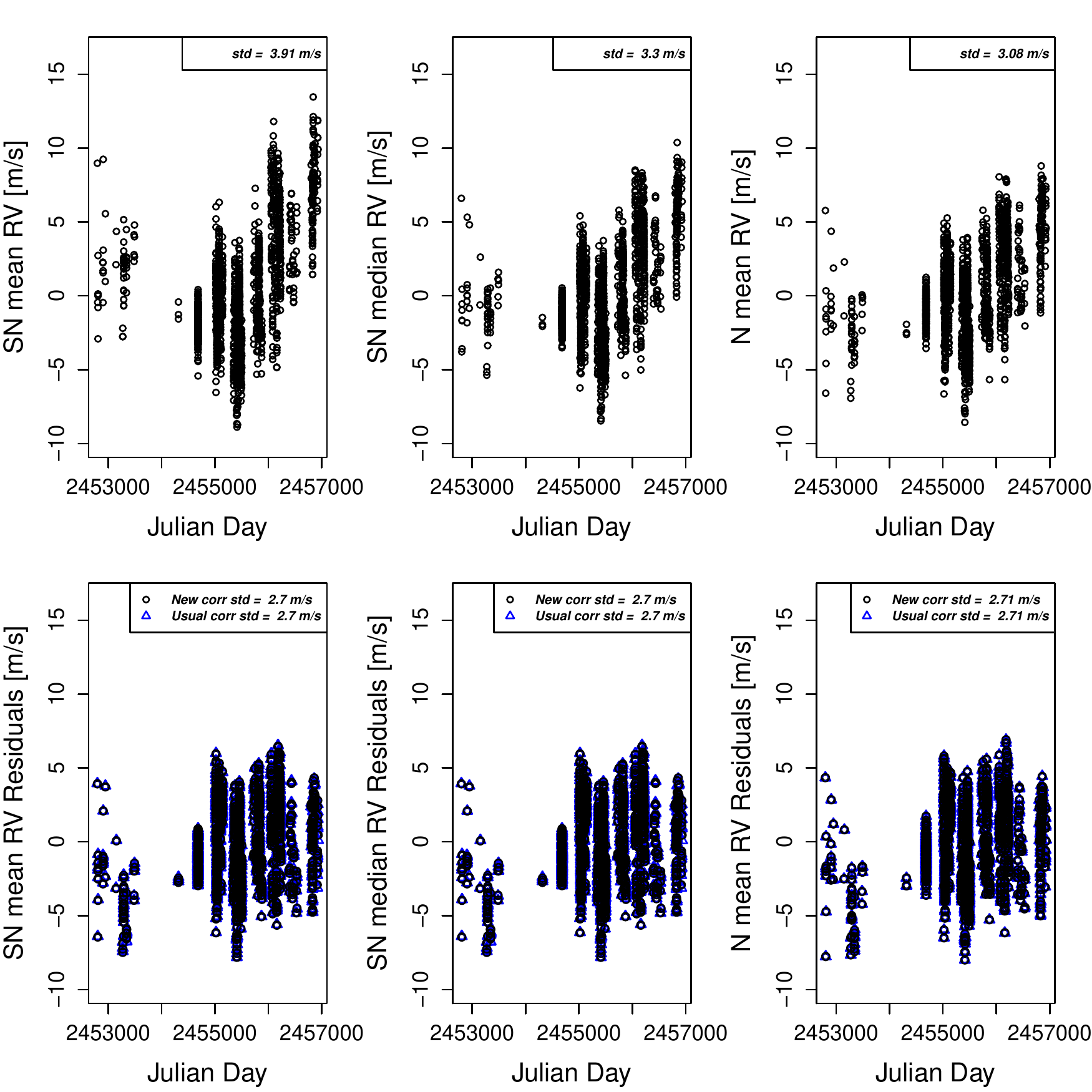} 
   \caption{(top) The RVs (black dots) for $\text{HD}192310$ estimated using a SN and a Normal fit.
 (bottom) The residuals from the model fit using Eq.~\eqref{eq:RV:correction} (New corr. std--black dots) and the residuals from the usual correction (Usual corr. std--blue triangles), based on $RV_{\text{activity}}=\beta_0+\beta_1 \gamma + \beta_2 \text{SN FWHM}$ for the SN fit and on $RV_{\text{activity}}=\beta_0+\beta_1 \text{BIS SPAN} + \beta_2 \text{FWHM}$ for the Normal fit. The residuals for both the proposed correction from stellar activity are comparable.
 }
   \label{fig:HD192310:correctionRV}
\end{center}
\end{figure*}

%HD10700
\begin{figure*}[htbp]
\begin{center}
\includegraphics[height = 6in]{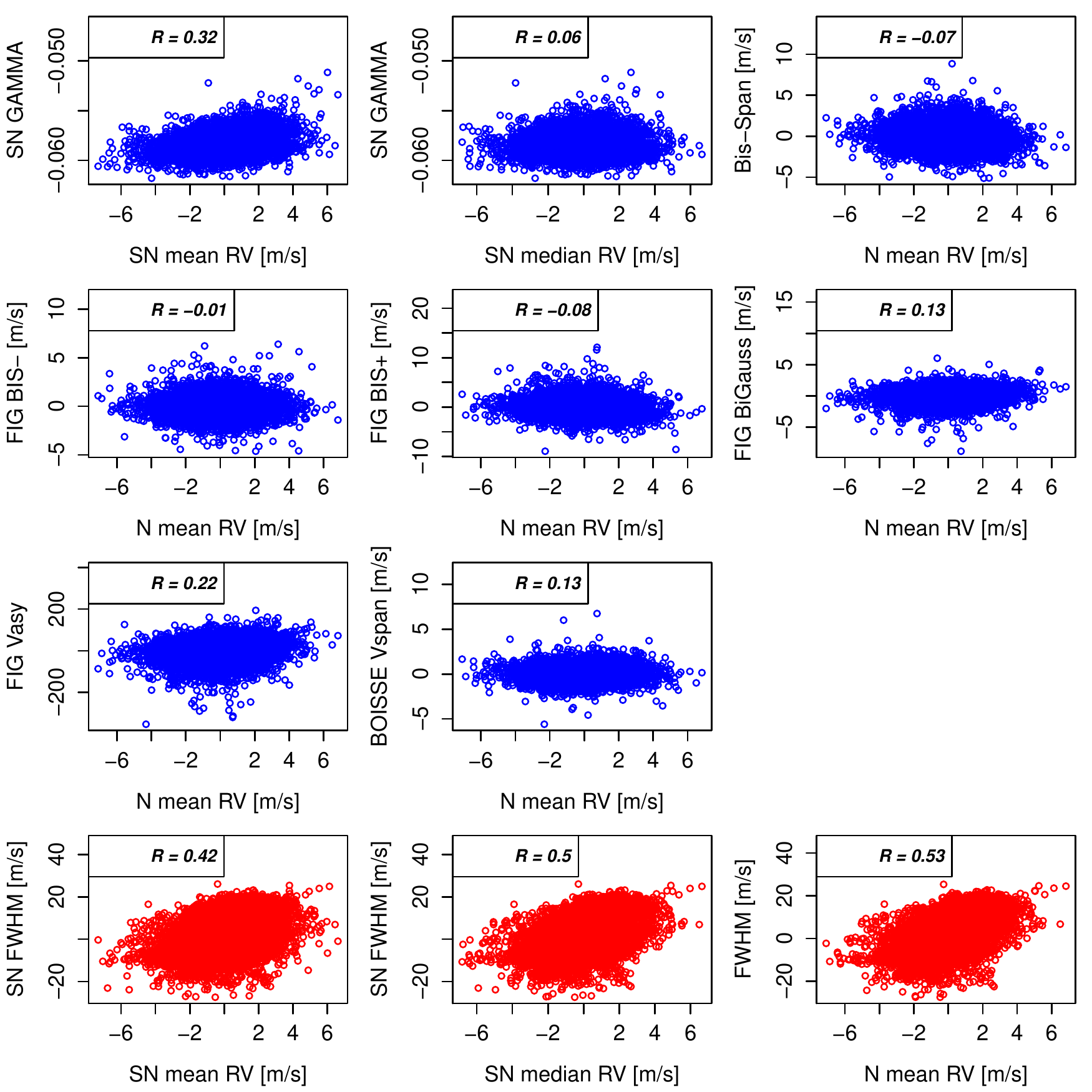}  
   \caption{(top three rows) Correlations between the asymmetry parameters and their corresponding RVs for $\text{HD}10700$. 
(bottom row) Correlations between the FWHM and the RVs for $\text{HD}10700$. 
The correlations are consistently stronger when using SN mean RV compared to N mean RV for the asymmetry parameters; however, the correlation between the FWHM and the N mean RV, only for this quiet star, is stronger the the analogous correlations with the estimated SN RVs. The estimated $R$ are statistically significant, except for the correlation between FIG BIS and RV (p--values=0.36).}   
   \label{fig:Tau:corrPlot}
\end{center}
\end{figure*}

\begin{figure*} 
\begin{center}
\includegraphics[height = 6in]{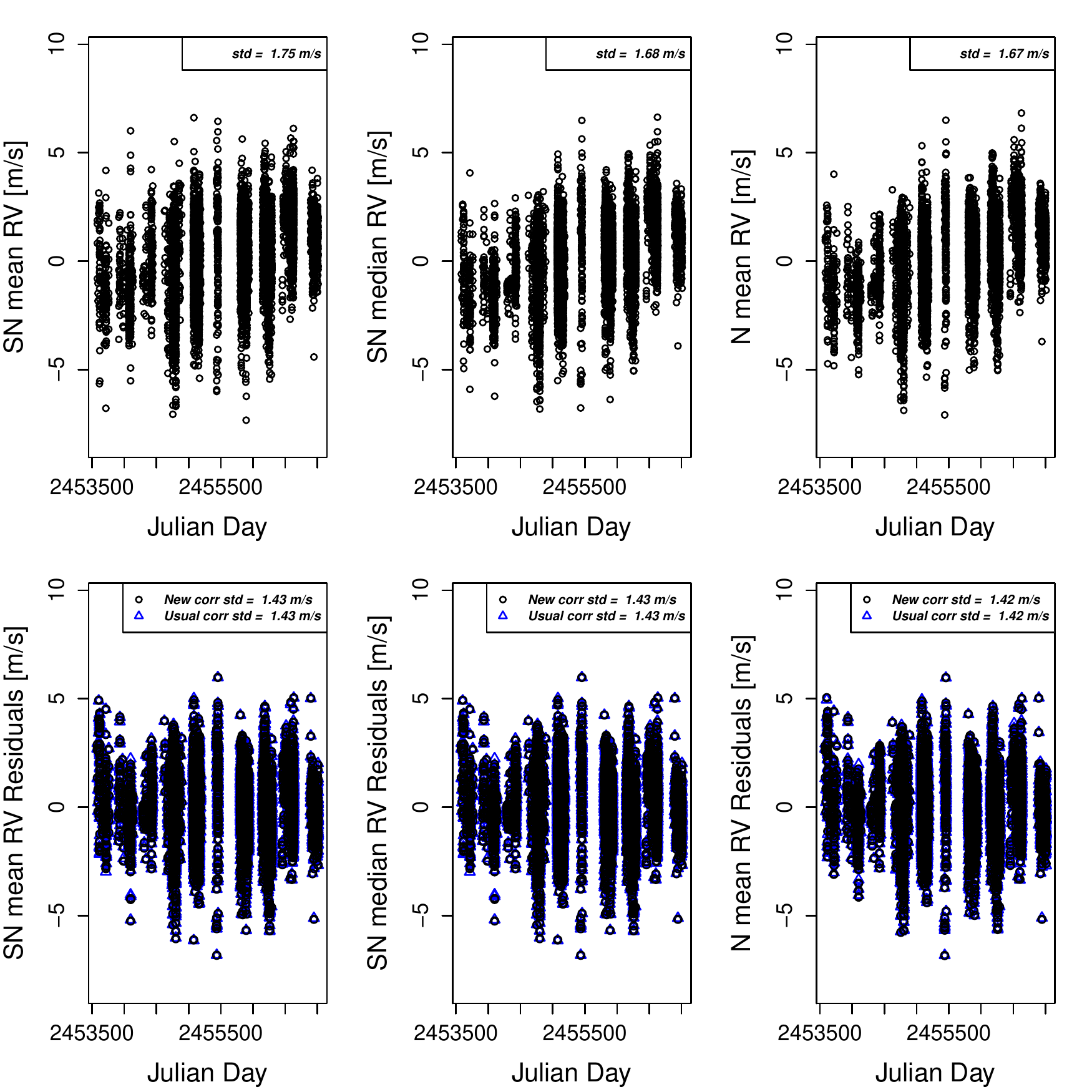} 
   \caption{(top) The RVs (black dots) for $\text{HD}10700$ estimated using a SN and a Normal fit.
 (bottom) The residuals from the model fit using Eq.~\eqref{eq:RV:correction} (New corr. std--black dots) and the residuals from the usual correction (Usual corr. std--blue triangles), based on $RV_{\text{activity}}=\beta_0+\beta_1 \gamma + \beta_2 \text{SN FWHM}$ for the SN fit and on $RV_{\text{activity}}=\beta_0+\beta_1 \text{BIS SPAN} + \beta_2 \text{FWHM}$ for the Normal fit. The residuals for both the proposed correction from stellar activity are comparable.}
   \label{fig:HD10700:correctionRV}
\end{center}
\end{figure*}

%HD215152
\begin{figure*}[htbp]
\begin{center}
\includegraphics[height = 6in]{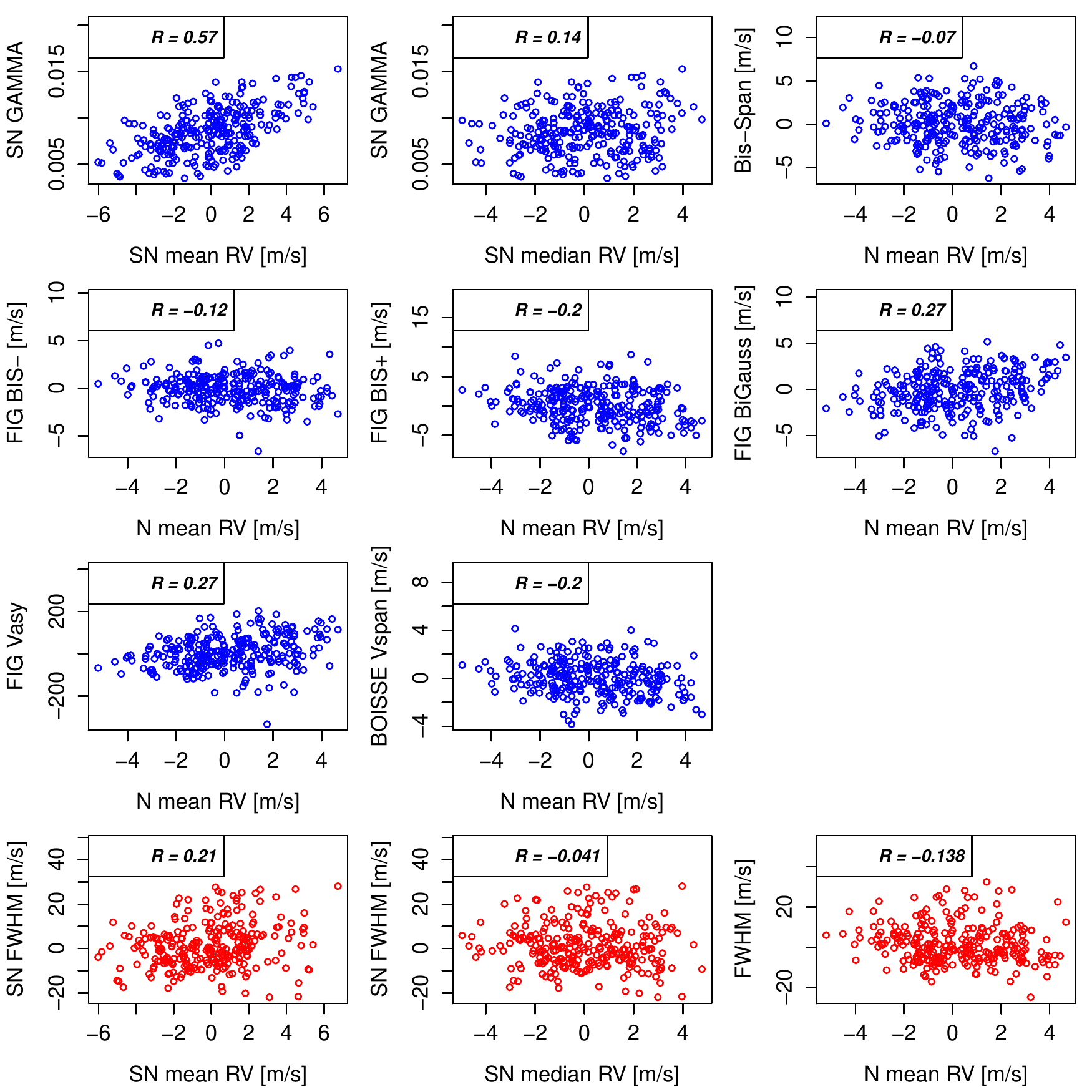}  
   \caption{(top three rows) Correlations between the asymmetry parameters and their corresponding RVs for $\text{HD}215152$. 
(bottom row) Correlations between the FWHM and the RVs for $\text{HD}215152$. 
The correlations are consistently stronger when using SN mean RV compared to N mean RV.
The p--values associated with each $R$ are not statistically significant for the correlation between N mean RV and BIS SPAN (p--values=0.27), the correlation between N mean RV and FIG BIS- (p--values=0.05),  the correlation between SN median RV and SN FWHM (p--values=0.5) and the correlation between N mean RV and FWHM (p--values=0.2).
}
   \label{fig:HD215152:corrPlot}
\end{center}
\end{figure*}

\begin{figure*} 
\begin{center}
\includegraphics[height = 6in]{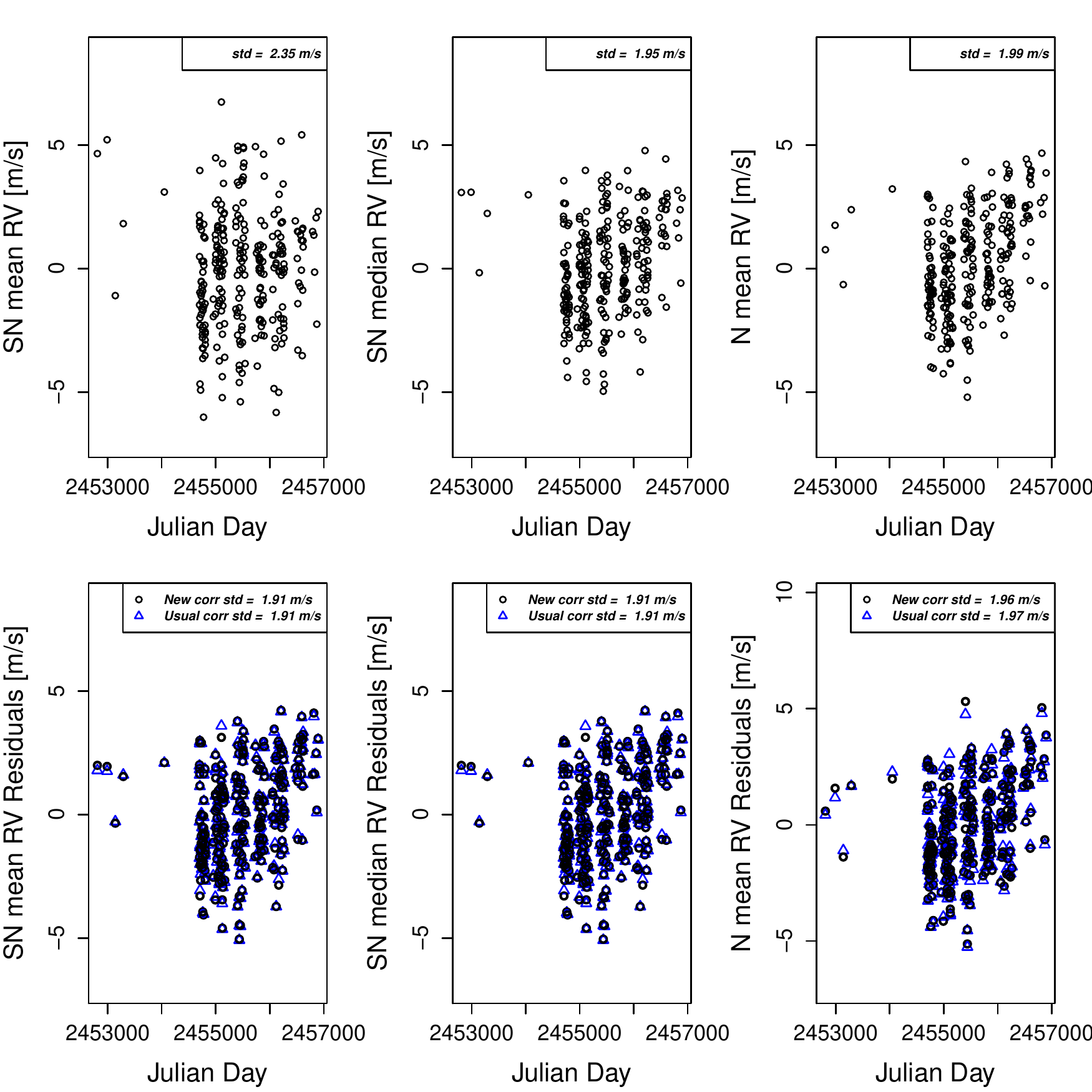} 
   \caption{(top) The RVs (black dots) for $\text{HD}215152$ estimated using a SN and a Normal fit.
 (bottom) The residuals from the model fit using Eq.~\eqref{eq:RV:correction} (New corr. std--black dots) and the residuals from the usual correction (Usual corr. std--blue triangles), based on $RV_{\text{activity}}=\beta_0+\beta_1 \gamma + \beta_2 \text{SN FWHM}$ for the SN fit and on $RV_{\text{activity}}=\beta_0+\beta_1 \text{BIS SPAN} + \beta_2 \text{FWHM}$ for the Normal fit. The residuals for both the proposed correction from stellar activity are comparable.}
   \label{fig:HD215152:correctionRV}
\end{center}
\end{figure*}

%Corot-7
\begin{figure*}[htbp]
\begin{center}
\includegraphics[height = 6in]{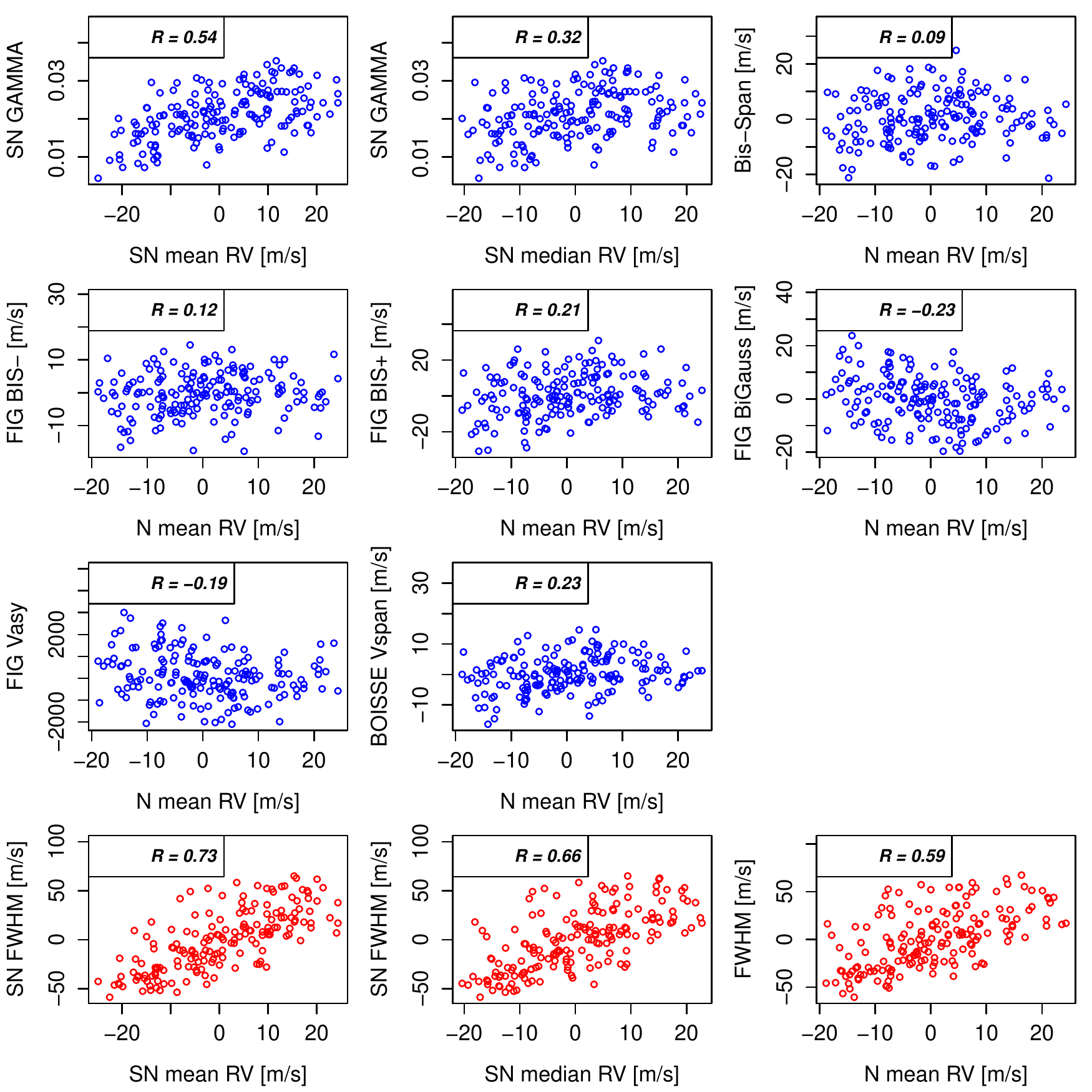} 
   \caption{(top three rows) Correlations between the asymmetry parameters and their corresponding RVs for $\text{Corot }7$. 
(bottom row) Correlations between the FWHM and the RVs for $\text{Corot }7$.
The correlations are consistently stronger when using parameters derived from the SN than the Normal.
The p--values associated with each $R$ are not statistically significant for the correlation between N mean RV and BIS SPAN (p--values=0.23) and the correlation between N mean RV and FIG BIS- (p--values=0.11).}
   \label{fig:Corot7:corrPlot}
\end{center}
\end{figure*}

\begin{figure*} 
\begin{center}
\includegraphics[height = 6in]{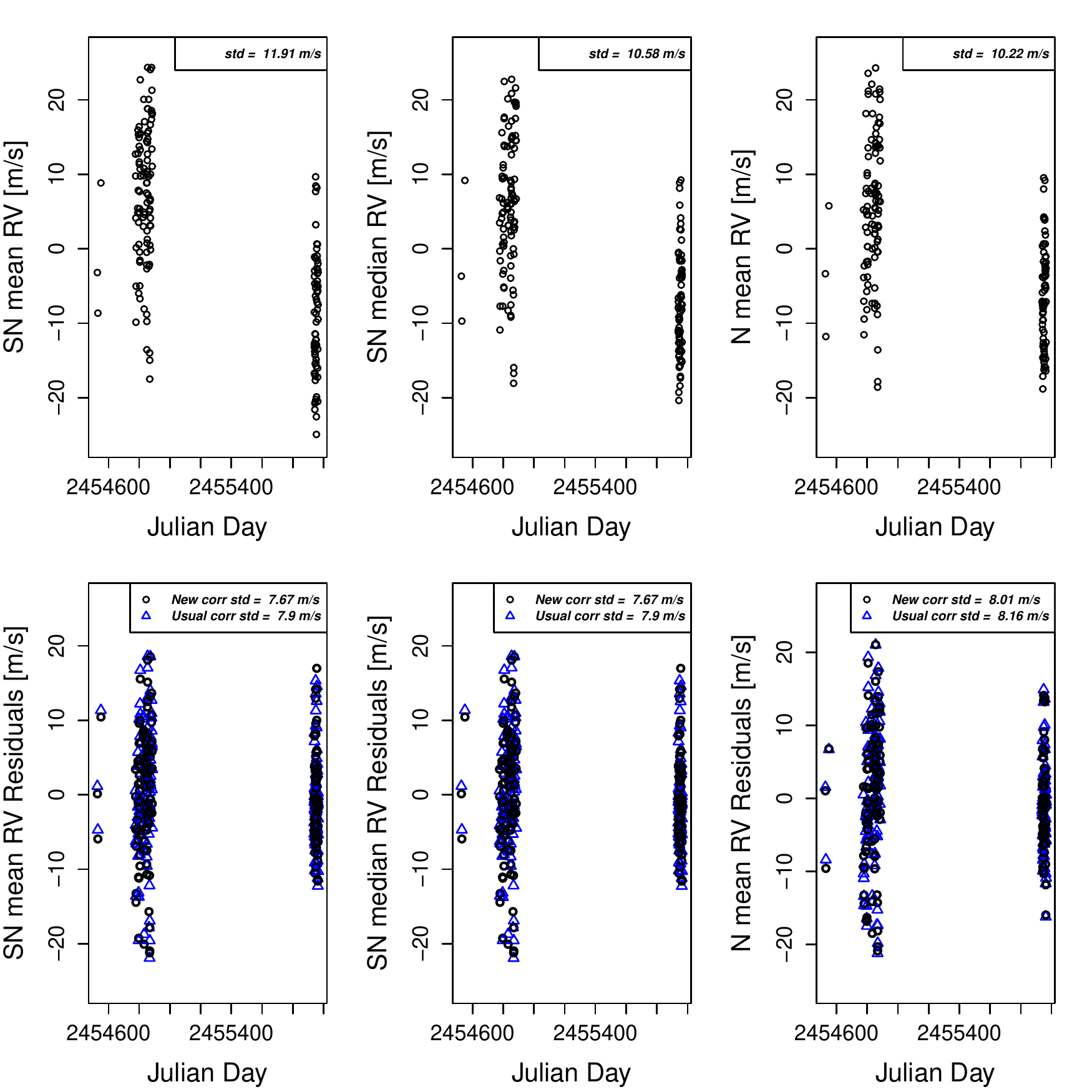} 
   \caption{(top) The RVs (black dots) for $\text{Corot }7$ estimated using a SN and a Normal fit.
 (bottom) The residuals from the model fit using Eq.~\eqref{eq:RV:correction} (New corr. std--black dots) and the residuals from the usual correction (Usual corr. std--blue triangles), based on $RV_{\text{activity}}=\beta_0+\beta_1 \gamma + \beta_2 \text{SN FWHM}$ for the SN fit and on $RV_{\text{activity}}=\beta_0+\beta_1 \text{BIS SPAN} + \beta_2 \text{FWHM}$ for the normal fit. The residuals have a smaller systematic component when using the proposed model of Eq.~\eqref{eq:RV:correction} (black dots) compared to the usual model (blue triangles). Moreover, once corrected for stellar activity using Eq. \ref{eq:RV:correction}, the remaining standard deviation from the SN models are $0.334$ \ms smaller than the remaining standard deviation of the Normal model.}
   \label{fig:Corot-7:correctionRV}
\end{center}
\end{figure*}

\bibliographystyle{aa}
%\bibliography{dumusque_bibliography}
\bibliography{mybib-SNCCF}

%\begin{appendix}
%\end{appendix}

\end{document}